\documentclass[twocolumn,superscriptaddress]{revtex4-2}

\usepackage{glossaries}

\newacronym{vqa}{VQA}{Variational Quantum Algorithms}
\newacronym{qa}{QA}{Quantum Annealing}
\newacronym{qaoa}{QAOA}{Quantum Approximate Optimisation Algorithm}
\newacronym{nisq}{NISQ}{Noisy Intermediate-scale Quantum}
\newacronym{sk}{SK}{Sherrington-Kirkpatrick}
\newacronym{ipr}{IPR}{Inverse Participation Ratio}
\newacronym{rqa}{RQA}{Reverse Quantum Annealing}
\newacronym{frio}{FRIO}{Fast Reverse Iterative Optimisation}

\usepackage{graphicx} 

\usepackage{epsfig,amsmath,amssymb,color,comment,physics, dsfont, bbold}
\usepackage[makeroom]{cancel}
\usepackage[caption=false]{subfig}
\usepackage{mathrsfs}
\usepackage[countmax]{subfloat}
\usepackage[normalem]{ulem}
\usepackage[english]{babel}
\usepackage{dsfont}
\usepackage{float}
\usepackage{color}
\usepackage[dvipsnames]{xcolor}
\usepackage{float}
\usepackage{url}
\usepackage{tikz}
\usepackage[bookmarks=true,colorlinks,linkcolor=blue,urlcolor=NavyBlue,citecolor=RoyalBlue]{hyperref}
\usepackage{cleveref}
\usepackage{bm}
\usepackage{enumerate}
\usepackage{booktabs}
\newtheorem{theorem}{Theorem}

\begin{document}
\title{Non-adiabatic Optimisation via Localisation–Delocalisation Transitions}

\author{Ryan Smith}
\affiliation{Phasecraft Ltd., Colston Street, Bristol BS1 4XE, United Kingdom}
\affiliation{School of Physics and Astronomy, University of Leeds, Leeds LS2 9JT, United Kingdom}
\author{Sabrina Yue Wang}
\affiliation{Phasecraft Ltd., Colston Street, Bristol BS1 4XE, United Kingdom}
\author{Filippo Maria Gambetta}
\affiliation{Phasecraft Ltd., Colston Street, Bristol BS1 4XE, United Kingdom}
\author{Raul A. Santos}
\email{raul@phasecraft.io}
\affiliation{Phasecraft Ltd., Charlotte Street, London W1T 4PW, United Kingdom}

\date{\today}

\begin{abstract}
The promise of efficiently solving classical optimisation problems on quantum computers is tantalizing. As many such problems can be mapped to approximating the ground state of spin-glasses, paradigmatic models for studying many-body localisation, it is natural to ask if some of the intuition coming from quantum statistical mechanics can help guide the construction of novel quantum optimisation algorithms.
In this work, we discuss how continuous reverse quantum annealing crossing a many-body localisation transition can be accelerated by considering the density of Landau-Zener crossings as the system is quenched.
We provide an interpretation of this behaviour in terms of an effective Fokker-Planck equation for the success probability of the algorithm.
This interpretation suggests an improvement in the required annealing time as a function of the system size $N$, which goes from $\mathcal{O}({\rm poly}(N))$ to $\mathcal{O}(1)$ for reference states with non-vanishing energy density. We show that the continuous dynamics can be Trotterised with a number of steps that remains modest as system size increases, making it appealing for near-term devices. For the resulting circuit implementation, we provide a quantitative strategy for determining the parameters of the time-dependent Hamiltonian. We compare the performance of our digitised algorithm, which we call \gls{frio}, against that of the \gls{qaoa} for finding the ground state of spin-glass problems, in particular the \acrlong{sk} model. The numerical results suggest that with suitably selected circuit parameters which can be efficiently found using thermalizing dynamics, \gls{frio} almost always outperforms the conventional \gls{qaoa} ansatz with pre-optimized angles. 

\end{abstract}

\maketitle

\makeatletter
\let\latex@addcontentsline\addcontentsline
\renewcommand{\addcontentsline}[3]{}
\makeatother

\section{Introduction}

Solving combinatorial optimisation problems on a quantum computer faster than classical methods is an active area of research, due to its immediate practical consequences in a range of applications, from managing power systems in energy grids to communications~\cite{ullah2022quantum,krikidis2026quantum} and finance~\cite{Herman2023Finance}. A popular class of methods to achieve this is that of \gls{vqa}~\cite{cerezo2021variational}, which are quantum-classical hybrid routines that aim to minimize the desired cost functions by turning the problem into finding the ground state of an appropriate Hamiltonian. A paradigmatic method is the \acrlong{qaoa}~\cite{farhi2014quantum}, which emerged as a natural candidate due to its scalability, flexibility, and suitability for near-term \gls{nisq} hardware platforms~\cite{blekos2024review}. 

Many optimisation problems can be reformulated as finding the ground states of disordered spin glass systems, often with high connectivity. Spin glasses are renowned for their extremely complex energy landscapes, riddled with local minima, and exhibit exotic behaviours stemming from randomness and frustration~\cite{dahlberg2025spin, sherrington202550}. Moreover, quantum spin glasses are paradigmatic models for studying many-body localisation (MBL) owing to the coexistence of strong disorder and interactions. As opposed to the usual ergodic phase, MBL phases are characterized by features such as local integrability, Poissonian level statistics and area-law eigenstate entanglement and logarithmic entanglement growth~\cite{Baldwin2016QREM,Mukherjee2018SKMBLtransition,lukin2025manybodylocalizationquantumising}.

\gls{qa}, based on continuous time evolution, is a widely used method for ground-state preparation in quantum systems, and has been extensively studied in the context of disordered systems~\cite{crosson2021prospects}. Traditional \gls{qa} requires annealing from the ground state of an easily prepared Hamiltonian. Since these are adiabatic protocols, they are naturally hard to implement in quantum devices due to the exponentially long annealing times required to maintain adiabaticity. This has prompted the development of \gls{rqa}~\cite{Chancellor2017Reverse,Yamashiro2019RA,Pelofske2023RA}, a technique which starts from a known low-energy classical bitstring. Along the annealing schedule, a transverse field is first turned on to increase quantum fluctuations promoting the exploration of nearby local minima, and is then turned off to get a potentially refined solution. While the process is non-adiabatic, it still requires long evolution times. Recently, other works have introduced a cyclic and iterative annealing protocol to reduce the annealing time to polynomial time by leveraging quench sequences around the MBL tricritical point~\cite{Wang2022MBLopt,Zhang2024annealing}.

\begin{figure*}
    \centering
    \includegraphics[width=0.99\linewidth]{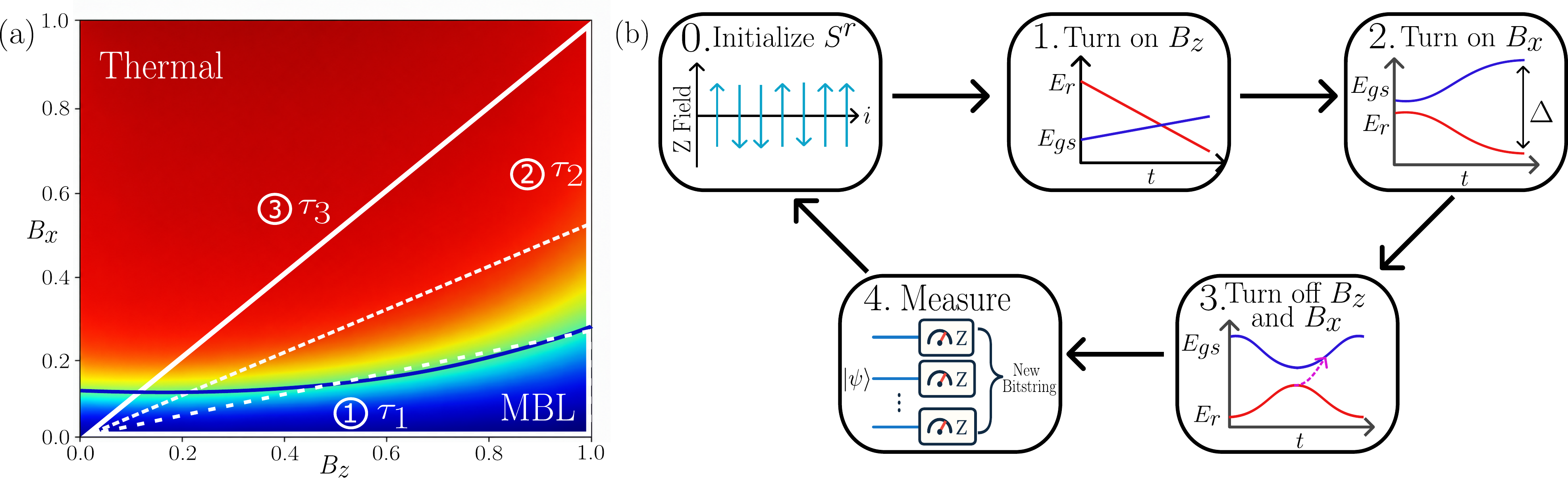}
    \caption{(a): Approximate phase diagram of the SK model of \cref{eq:PhasediagramHamiltonian} for small system sizes achievable using exact diagonalization (ED) for $N=12$, showing the first three steps of the quench protocol for $\chi = 0.25$ (dotted), $\chi=0.5$ (dashed), and $\chi = 1$ (solid). The blue line is the approximate  critical line that separates the many-body localized phase from the thermal delocalized phase that the cyclic algorithm  must traverse to be successful. (b): Pictorial demonstration of the steps of FRIO. We initialise the direction of the longitudinal fields according to the reference state, followed by a series of three quenches which traverses the triangular path of the phase diagram according to chosen $\chi$ and finally an $N$-qubit measurement in the computational basis.}
    \label{fig:phase_diagram}
\end{figure*}

In this work, we combine elements from several types of \gls{qa} protocols to design an optimisation algorithm for obtaining low energy states of spin-glasses which is more resilient to non-adiabaticity, Trotterisation, and is amenable to implementation on NISQ devices. We call it Fast Reverse Iterative Optimisation or FRIO. Part of FRIO's resilience comes from exploring the localisation-delocalisation transition which controls the tunneling between local minima when a transverse field is turned on. The efficacy of this method is tested on the problem of finding the ground state of disordered spin-glasses, specifically the \gls{sk} model and weighted $k$-local graphs. The paper is organized as follows. In section \ref{sec:methods} we describe FRIO, the differences with the algorithm described in ref.~\cite{Wang2022MBLopt}, and the intuition behind FRIO's improved parameters based on an effective hopping model between energy levels and a coarse-grained description of the success probability in terms of a Fokker-Planck equation. We also present a method to determine the protocol's parameters needed for successful cooling by inspecting short-time thermalization dynamics. 
In \cref{sec:results}, we demonstrate numerically that our thermalization approach can be directly implemented to determine useful circuit parameters for FRIO. We then compare the efficacy of our algorithm against that of an optimized \gls{qaoa} protocol. We conclude in \cref{sec:conclusion} and discuss further research directions.

\section{\label{sec:methods} Continuous-time Annealing}

\subsection{Adiabatic reverse annealing using many-body localisation}
In this work, we mainly focus on instances of the \gls{sk} model, a paradigmatic model in condensed matter physics that describes a spin glass with infinite-range random interactions~\cite{Sherrington1975SKmodel}. The Hamiltonian is given by

\begin{equation}\label{eq:SKHam}
    H_{\rm SK}=\sum_{1\leq i<j\leq N} J_{ij}Z_i Z_j,
\end{equation}
where $Z_i$ are the Pauli-$Z$ matrices that represent binary variables $i=1,2,...N$ and $J_{ij}$ is the cost matrix whose matrix elements are independently sampled from a normal distribution with zero mean and variance $J^2/N$, $\mathcal{N}(0,J^2/N)$. For ease, we fix $J=1$. All $2^N$ eigenstates of \cref{eq:SKHam} are encoded by bitstrings $b_n=(s_1^{n_1},\dots,s_N^{n_N})$, $s^{n_i}_{i}=\pm1$, with corresponding eigenenergies $E_{\alpha}=\sum_{1\leq i<j\leq N}s_i^{n_i}J_{ij}s_j^{n_j}$ that form a glassy energy landscape with exponentially many local minima~\cite{Edwards1975spinglass,Mezard1986,Crisanti2003SK}.

In adiabatic \gls{qa}, one typical strategy to find the ground state of spin glass Hamiltonians is to add time-dependent, non-commutative terms to the system Hamiltonian (the optimisation cost function). The most common of these ``quantum" terms is a time-dependent transverse magnetic field applied in the $x$-direction,
\begin{equation}
    \begin{array}{cc}
        H(t)=H_{\rm SK}+B_x(t)H_x, & H_x=-\sum_{i=1}^N X_i.
    \end{array}
\end{equation}
The qubits are then initialised in the $x$-polarized state $|\psi_0\rangle=|+\rangle^{\otimes N}$, which for $B_x\gg J$  is the effective ground state of $H_x$ and is separated by a large gap from the rest of the spectrum \cite{Laumann2014annealing,Baldwin2017annealing}. If $B_x(t)$ is adiabatically turned off, then the system should not undergo any Landau-Zener transitions \cite{zener1932non,Landau1932PZS2} and the state should follow the instantaneous ground state towards the minimum of the  \gls{sk} energy landscape. However, since the required annealing rate for this to happen is $\tau^{-1}_{\rm anneal}\ll\Delta^{2}/B_x$, where $\Delta$ is the minimal avoided crossing gap, this in general requires exponentially long annealing times as $\Delta$ can decay exponentially with the system size in the worst case, i.e., $\Delta \sim e^{-cN}$~\cite{Amin2009AQC,Altshuler2010AQOfail,Jörg2010QAfail}. 

To overcome this challenge, recent works~\cite{Wang2022MBLopt,Zhang2024annealing} have designed a cyclic reverse annealing algorithm for optimisation that iteratively passes through local energy minima by circling the tricritical point located between the spin-glass, MBL, and thermal phases. We adopt a similar protocol for FRIO, but with some important differences like much shorter annealing times than those required in \cite{Wang2022MBLopt}. 
Our algorithm consists of four steps discussed below, which are summarized in Fig. \ref{fig:phase_diagram}(b).

{\em Step 1.} Starting from a reference bitstring $S^r$ with energy $E_r$ that serves as the initial state, apply a longitudinal magnetic field that has $S^r$ as its minimum energy state. The system is then quenched as
\begin{equation}\label{eq:H_1}
\begin{array}{cc}
    H_1(t)=H_{\rm SK}+B_z(t)H_z, & H_z=-\sum_{i=1}^N S^r_iZ_i,
\end{array}
\end{equation}
where $B_z(t)=B_zt$ is a linear ramp function. This makes $S^r$ the new ground state when $B_z>B_z^c$, where $B_z^c=(E_r-E_{GS})/N$ is the critical field strength since there is zero overlap. Clearly this step is purely classical and no Landau-Zener transitions can occur. Hence, the final state does not depend on the speed of the quench so we restrict the quench time to $t\in[0,1]$.

{\em Step 2.} Perform a second quench by adding a transverse magnetic field to the system Hamiltonian,
\begin{equation}\label{eq:H_2}
\begin{array}{cc}
    H_2(t)=H_{\rm SK}+B_zH_z+B_x(t)H_x,
\end{array}
\end{equation}
where $B_x(t)= t$ is another linear ramp. The annealing time here is $t\in[0,\chi B_z]$, with $\chi=B_x/B_z$, so that at the end of the ramp $B_x(\chi B_z) = B_x$. We increase $\chi=B_x/B_z$ until it reaches the desired ratio which determines how far we quench (if at all) into the thermal phase. The value of $B_x=\chi B_z$ determines the size of the Landau-Zener transition gap~\cite{Landau1932PZS2,zener1932non}. This step only requires $\mathcal{O}(1)$ time to increase $B_x$ due to the gap being proportional to the total magnetic field strength and independent of system size. 

\begin{figure*}
    \centering
    \includegraphics[width=0.99\linewidth]{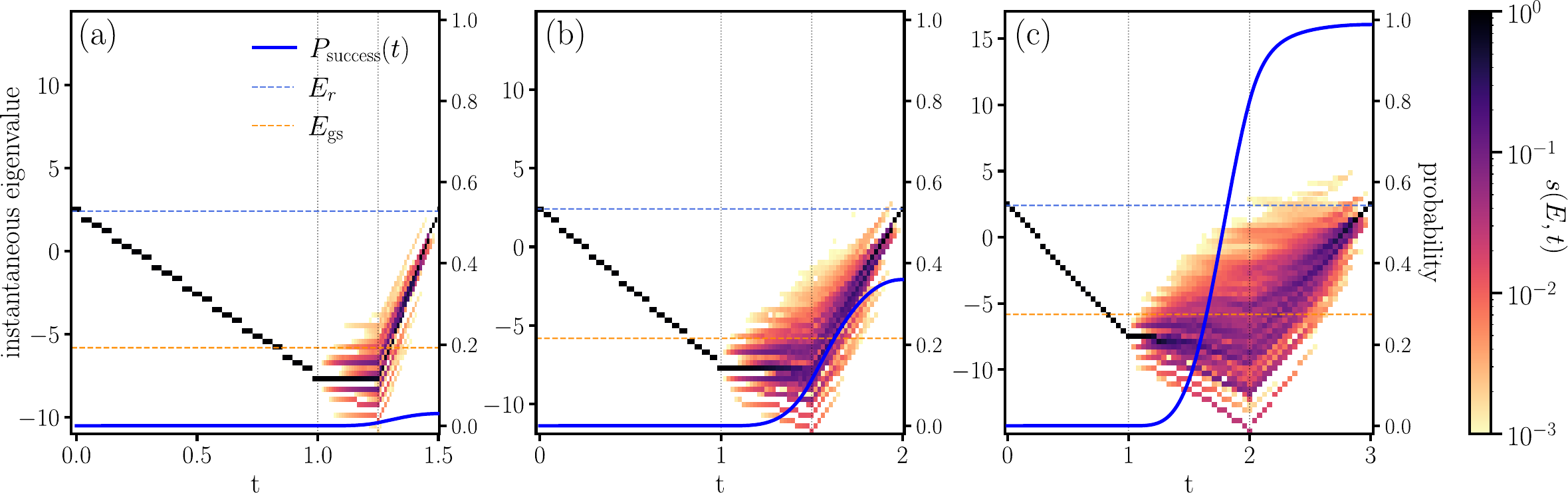}
    \caption{Evolution of the spectral weight $s(E,t)=\sum_n|\langle n(t)|S^r\rangle|^2\delta(E-E_n(t))$ in the FRIO algorithm for a system with 10 sites, showing the first three steps of the quench protocol for $\chi = 0.25$, $0.5$, and $1$ (left to right). Here $|n(t)\rangle$ is the instantaneous eigenvector with instantaneous eigenvalue $E_n(t)$. In blue we plot the probability of a successful energy transition, $P_{\rm success}$, during the ramping sequence associated with each value of $\chi$. The dashed blue line marks the energy of the initial state, while the orange dashed line marks the ground-state of $H_{\rm SK}$. The vertical grey dashed lines mark the beginning of the second and third step of the algorithm.}
    \label{fig:spectral}
\end{figure*}

{\em Step 3.} Perform a third quench back to the \gls{sk} model using the Hamiltonian
\begin{equation}\label{eq:H_3}
    H_3(t)=H_{\rm SK}+B^{(3)}_{z}(t)H_z+B_{x}^{(3)}(t)H_x,
\end{equation}
where $B^{(3)}_x(t)=\chi B^{(3)}_z(t)$ and $B_z^{(3)}(t)= B_z -t/\chi$ is the final linear ramp. Here the annealing time is $t\in [0,\chi B_z]$. The system then crosses the phase boundary between the paramagnetic and spin glass phases. Landau-Zener transitions that occur can be tuned by changing $\chi$ to increase the probability of landing on a state of lower energy than $E_r$~\cite{Wang2022MBLopt}.

{\em Step 4.} Measure in the computational basis. If the energy $E_{\rm meas}$ of the measured bitstring $S^{\rm meas}$ is less than $E_r$, then the cycle can be repeated with $S^{\rm meas}$ being the new reference state. Otherwise, the protocol is repeated with $S^r$, possibly changing the quench rates. 

The total annealing time of this protocol is $t_{ f}=\tau_1+\tau_2+\tau_3=1+ 2\chi B_z$ where $\tau_1,\tau_2,\tau_3$ are the annealing times for each quench in the sequence. In Ref.~\cite{Wang2022MBLopt}, it was proposed that approaching the tricritical point from above, the gap of the avoided crossings scales inverse polynomially with the system size. Imposing an annealing time $\sim \chi B_z= \mathcal{O}({\rm poly}(N))$, the protocol should avoid diabatic transitions and lead to cooling. 

\subsection{Non-adiabatic cyclic quantum annealing}
\label{subsec:nonadiabatic}

The main source of the $\mathcal{O}({\rm poly}(N))$ annealing time in \cite{Wang2022MBLopt} is the requirement of avoiding Landau-Zener transitions during step 3.\ of the protocol above. This requirement, while guaranteeing the success of the algorithm, is quite restrictive, especially for implementation in near-term quantum computers. In this work, we show that this requirement can be relaxed by considering instead an algorithm that returns a non-zero success probability, defined as
\begin{align}
\label{eq:succ_prob}
    P(E(t_f)< E_r)=\sum_{n: E_n(t)< E_r}|\langle b_n(t_{f})|S^r\rangle|^2, 
\end{align}
where $|b_n(t)\rangle =U(t)|b_n\rangle$ is the instantaneous eigenstate at time $t$ and  $U(t)=\mathcal{T}e^{-i\int_0^{t}H(t')dt'}$, even in the presence of a shorter annealing time. We do this by computing semiclassically the success probability and showing that, under very natural assumptions, it is possible to achieve a probability larger than 1/2 even if the annealing drive is not strictly adiabatic. Our algorithm can still find good approximations to the ground-state energy even for annealing times as short as $t_f =\mathcal{O}(1)$.

To start, we can gain intuition on the performance of this algorithm by studying the propagation of the spectral weight during the quench, solving the time-dependent Schr\"{o}dinger equation (TDSE). This is shown in \cref{fig:spectral}. The first part of the quench just pushes the initial state to the bottom of the spectrum. The second step allows tunneling to nearby states. Some of these nearby states can be high energy with respect to the original target Hamiltonian $H_{\rm SK}$, so a final quench is needed to return to the original Hamiltonian. As a comparison we also plot the probability of a successful transition to a lower energy state upon performing a measurement, \cref{eq:succ_prob}, during the non-adiabatic quench protocol. By using larger $\chi$, the reference state $S^r$ can more effectively tunnel to nearby states, directly increasing the probability of measuring a bitstring that lies in a lower energy minimum. While our analysis is restricted to a reference state corresponding to a mid-spectrum eigenstate, we show in \cref{sec:hybrid} that it remains valid for low-lying excited eigenstates, where the rate of transitions to lower-energy states is marginally suppressed \cite{SOM}.

Motivated by \cref{fig:spectral}, we look for an average description on an energy window much larger than the typical level-spacing, such that a coarse grained approach is reasonable. Assuming that, at a given crossing, the gap $\Delta$ between two energy levels is distributed according to the Wigner's surmise \cite{mehta2004randommatrices}
\begin{align}
P(\Delta,\bar{\Delta})=\frac{\pi\Delta}{2\bar{\Delta}^{2}}e^{-\pi\frac{\Delta^2}{4\bar{\Delta}^2}},
\end{align}
where $\bar{\Delta}$ is the mean level spacing between the energy levels that cross the reference state (not the mean level spacing of all the levels), then the average Landau-Zener probability for a given mean level spacing $\bar{\Delta}$ is the average of the adiabatic probability $p_{\rm nLZ}(\Delta)=1-e^{-c\frac{\Delta^2}{v}}$ over all the available levels,
\begin{align}
    \langle p_{\rm nLZ}\rangle_{\bar{\Delta}}=\int_0^\infty P(\Delta,\bar\Delta)p_{\rm nLZ}(\Delta)d\Delta=\frac{4c\bar\Delta^2/(\pi v)}{1+4c\bar\Delta^2/(\pi v)},
\end{align}
where $c=\mathcal{O}(1)$ and $v=|dE/dt|$, which we assume to be constant over all transitions \footnote{We could consider a distribution over different velocities, but this technicality is not very illuminating for this discussion.}. Here we distinguish two limiting cases. For $\bar\Delta^2/v\gg1$, $\langle p_{\rm nLZ}\rangle_{\bar\Delta}\rightarrow 1$ and we recover the usual adiabatic limit. In the opposite diabatic limit $\bar\Delta^2/v\ll1$ we have instead $\langle p_{\rm nLZ}\rangle_{\bar\Delta}\sim 4c\bar\Delta^2/(\pi v) \ll 1 $.

Using the average Landau-Zener probability, we model the energy level transitions as a Poisson process, where the underlying assumption is that each level crossing is independent. At each avoided crossing the system follows the adiabatic level with a probability $\propto \langle p_{\rm nLZ}\rangle_{\bar\Delta}$. Under this assumption and the Kramers-Moyal expansion \cite{Kramers_1940,Moyal_1949} we can write down a master equation for the probability of observing an energy $E+dE$ at time $t$ as
\begin{equation}
   \partial_t P(E,t) = -\partial_E\big[A^{(1)}P\big] + \frac{1}{2}\,\partial_E^2\big[A^{(2)}P\big],
\end{equation}
where we have discarded higher energy derivatives (see derivation in \cref{app:Master_eq} of the Supplemental Material (SM)~\cite{SOM}). Here the drift and diffusion coefficients $A^{(1)}$ and $A^{(2)}$ are determined by the compound Poisson process and given by (see discussion in \cref{app:rates_Poisson} of the SM~\cite{SOM})
\begin{equation}
  A^{(1)} \sim\; \rho\,\langle p_{\rm nLZ}\rangle\,\bar\Delta\,\eta,
  \quad
  A^{(2)} \sim\; \rho\,\langle p_{\rm nLZ}\rangle\,\bar\Delta^2,
  \label{eq:A1A2}
\end{equation}
where $\rho$ is the density of states available to cross the reference state (and, in principle, independent of $\bar\Delta$) and $\eta$ is the asymmetry parameter
\begin{align}
    \label{eq:asym_parameter}
    \eta=\frac{N(E<E_r)-N(E>E_r)}{N(E<E_r)+N(E>E_r)},
\end{align}
which, for a given crossing at energy $E$, counts the normalized difference of states above and below the reference energy on a region of states connected to the reference state through the dynamics.
Treating $A^{(1)}$ and $ A^{(2)}$ as approximately constant over the sweep (which is justified in the vicinity of the tricritical slope $\chi\to\chi_{\rm tri}^+$, with $\chi_{\rm tri} \approx 3.6$ ~\cite{Wang2022MBLopt}, where the local crossing statistics vary slowly), the drift--diffusion equation with initial condition
$P(E,0)=\delta(E-E_r)$ has the exact solution
\begin{equation}\label{eq:Sol_FP}
  P(E,t) = \frac{1}{\sqrt{2\pi A^{(2)} t}}
  \exp\!\left[-\frac{\big(E - E_r - A^{(1)} t\big)^2}{2A^{(2)} t}\right].
\end{equation}
The success probability is obtained by integrating \cref{eq:Sol_FP} for $t=t_f$ over the energy,
\begin{equation}
  P(E(t_f) < E_r)
  = \int_{-\infty}^{E_r} P(E,t_f)\,dE
  = \frac{1}{2}\,{\rm erfc}(-x),
  \label{eq:success}
\end{equation}
with $\rm erfc(x)$ the complementary error function and $x$ given by 
\begin{equation}
  x = 
  \sqrt{t_f}\,\frac{A^{(1)}}{\sqrt{2A^{(2)}}}\sim\sqrt{t_f \rho\langle p_{\rm nLZ}\rangle}\eta,
  \label{eq:SNR}
\end{equation}
where we have used \cref{eq:A1A2}.
The limiting cases $|x|\gg1$ and $|x|\ll1$ can be found directly from the properties of the complementary error function
\begin{align}\label{eq:limits}
\centering
\begin{array}{lll}
  x\gg 1 & \Rightarrow & P(E(t_f)\leq E_r)= 1 -\mathcal{O}(\frac{e^{-x^{2}}}{2x\sqrt{\pi}}), \\
  |x|\ll 1        & \Rightarrow & P(E(t_f)\leq E_r)=\frac{1}{2}+\mathcal{O}(\frac{x}{\sqrt{\pi}}), \\
  x\ll -1 & \Rightarrow & P(E(t_f)\leq E_r)=\mathcal{O}(\frac{e^{-x^{2}}}{2|x|\sqrt{\pi}}).
\end{array}
\end{align}
Crucially, any $x>0$ produces a success probability larger than 1/2.
In the limit of fast annealing and using the expected density of states for spin glass problems \cite{Cavagna_2004}, $\rho\sim e^{\alpha N}$, together with the behaviour of the average gap $\bar\Delta\sim N^{-z/d_{\rm eff}}$ \cite{Koh2016SKexcitations}, $x$ depends on the system size $N$ as
\begin{equation}\label{eq:scaling}
    x\sim \eta\sqrt{t_f} e^{\alpha N/2}N^{-z/d_{\rm eff}},
\end{equation}
which even for $t_f=\mathcal{O}(1)$ grows with the system size (for $\eta>0$) and hence a growing success probability. This is the main insight that motivates the exploration of fixed (i.e., independent of the system size) annealing protocols. Clearly, \cref{eq:scaling} rests on a series of assumptions that can be broken. We discuss these assumptions in \cref{sec:validity} of the Supplemental  Material. 
 The differences between our proposed FRIO algorithm and the algorithm introduced in Ref.~\cite{Wang2022MBLopt} are captured in Table \ref{tab:comp_algos}.
\begin{table}[]
    \caption{Annealing times as a function of the system size $N$ for our protocol and the one discussed in Ref.~\cite{Wang2022MBLopt}. By considering a coarse grained description of the Landau-Zener transitions, we find that a constant annealing time can produce a favourable success probability. Step 1, being purely classical, does not have a natural annealing time.}
    \centering
    \begin{tabular}{l c c c}
        \toprule
        \textbf{Reference} & \textbf{Step 1} & \textbf{Step 2} & \textbf{Step 3} \\
        \midrule
        Ref.~\cite{Wang2022MBLopt} & --- & $\mathcal{O}(1)$ & $\mathcal{O}(\mathrm{poly}(N))$ \\
        This work                  & --- & $\mathcal{O}(1)$ & $\mathcal{O}(1)$ \\
        \bottomrule
    \end{tabular}
    \label{tab:comp_algos}
\end{table}

This analysis also signals where we expect FRIO to fail. As we see from \cref{eq:SNR} and \cref{eq:limits}, as we repeat the algorithm and successfully lower the energy, the fraction of states above and below the reference energy in a given energy window can change, eventually making $\eta$ negative, thus reducing the success probability below 1/2. As we approach the bottom of the spectrum this is expected as the density of states goes from exponential to polynomial to constant, recovering the usual obstruction produced by adiabaticity. 

Determining the optimal value of $\eta$ is nontrivial in practice, since it depends on an energy window over which the coarse-grained description is applicable. Nonetheless, this section serves as a motivation for the study of non-adiabatic annealing protocols. As we will show in \cref{sec:results} below, the parameters obtained by our numerical simulations indicate that even annealing protocols with $\chi<\chi_{\rm tri}$ can still give rise to successful runs even at low energy densities, suggesting that approaching the tricritical point may not be necessary.

The rest of this work focuses on understanding the effective performance of FRIO and, in particular, to assess its digitised Trotterised version in comparison with QAOA with similar circuit resources.

\subsection{digitised non-adiabatic cyclic quantum annealing}\label{subsec:Trotter 1}

Motivated by the possibility of a protocol with fixed, size-independent annealing time and the current constraints in digital quantum simulation, we Trotterise this protocol and analyze its behaviour with a small number of Trotter steps. The evolution unitary is 
\begin{align}\label{eq:annealing_path}
    U_\triangle(t)=\begin{cases}
        U_1(t) \quad &\mbox{for $t\in[0,t_1]$}\\
        U_2(t-t_1)U_1(t_1) \quad &\mbox{for $t\in[t_1,t_2]$}\\
        U_3(t-t_2)U_2(t_2-t_1)U_1(t_1)\quad  &\mbox{for $t\in[t_2, t_3]$}
    \end{cases}
\end{align}
where $U_i(t)=\mathcal{T}e^{-i\int_0^{t}H_i(t')dt'}$ is the time ordered evolution and 
$t_1=1$, $t_2=t_1+\chi B_z$ and $t_3=t_2+\chi B_z$. We proceed to approximate each $U_i(t)$ by $U_i^{\rm Tr}(t)$
with
\begin{align}\label{eq:first_order_Trotter_decomp}
    \centering 
        U^{\rm Tr}_1(t)&=e^{-iH_{\rm SK} t} e^{-iH_z \int_{0}^{t}B_{z}(t')dt' },\\
        U^{\rm Tr}_2(t)&=\prod_{k=1}^{M}(e^{-\frac{it}{M}(H_{\rm SK}+B_zH_z)} e^{-\frac{it}{M}B_x(k \frac{t}{M} )H_x}),\\
        U^{\rm Tr}_3(t)&=\prod_{k=1}^{M}(e^{-\frac{it}{M}(H_{\rm SK}+B_z^{(3)}(k \frac{t}{M})H_z)} e^{-\frac{it}{M}B_x^{(3)}(k \frac{t}{M})H_x}),
\end{align}
which are a first-order Trotter approximation of the evolution with $M$ layers (for $U_{2,3}$). We note that since the quench corresponding to $U_1(t)$ is entirely classical this step can be performed exactly using a single Trotter step. In fact, since the initial reference state is just a bit string, we can ignore the effect of the first quench as this would just apply a global phase.

We can compare directly the results of FRIO with \gls{qaoa} by choosing $p=3M$, where $p$ is the \gls{qaoa} depth. As an example that we later consider in Sec.~\ref{sec:results}, our cyclic algorithm with $M=1$ Trotter step is directly comparable to a \gls{qaoa} circuit of depth $p=3$. For an SK instance of size $N$, \gls{qaoa} and the cyclic algorithm obey an $\mathcal{O}(N^2)$ scaling in the number of two-qubit gates required to execute the circuit assuming all-to-all connectivity. The actual scaling for all the considered algorithms is summarized in Table~\ref{table:comp table}. In addition, we also consider a second-order Trotterisation of FRIO, which we dub $\mathrm{FRIO}_2$, that we outline in~\cref{sec:Trotter 2}.

\begin{table} 
    \caption{Comparison between FRIO and \gls{qaoa} for the \gls{sk} model in the number of one- and two-qubit rotation gates required to implement each algorithm. In this work, $M$ is the number of Trotter steps used in digitised FRIO. We define the rotation gates as $e^{-i\theta_x X_i}$, $e^{-i\theta_z Z_i}$, and $e^{-i\theta_{zz} Z_iZ_j}$, where $X,Z$ are the Pauli matrices.}
\centering
\begin{tabular}{l c c c}
        \toprule
        \textbf{Algorithm} & \textbf{\# Layers} & \textbf{\#1Q gates} & \textbf{\#2Q gates} \\
        \midrule
        FRIO & $6M$ & $4MN$ & $MN(N-1)$ \\
        QAOA              & $6M$ & $3MN$ & $\frac{3}{2}MN(N-1)$ \\
        \bottomrule
    \end{tabular}
\label{table:comp table}
\end{table}

\subsection{Parameter selection via thermalization}\label{subsec:parameter_scaling}

\begin{figure*}
    \centering
    \includegraphics[width=0.99\linewidth]{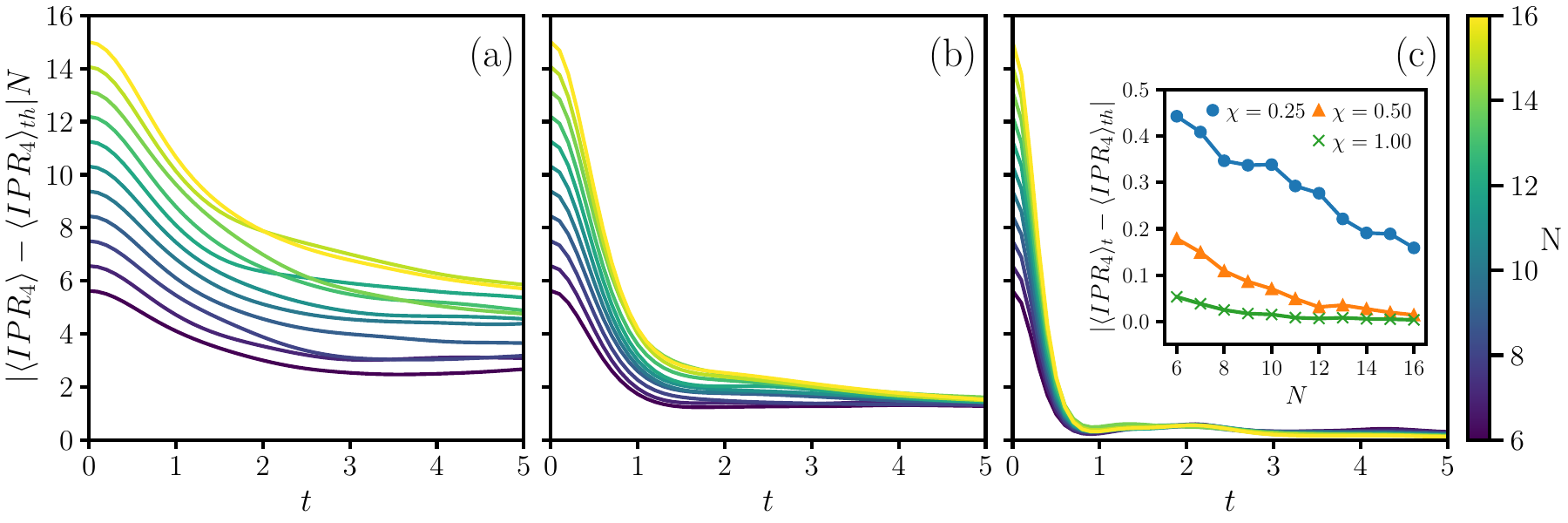}
    \caption{Marginal \gls{ipr} dynamics for parameter testing experiments using \gls{sk} type graphs. (a): The short-time dynamics of the rescaled \gls{ipr} for a $4$-qubit marginal performed at $\chi=0.25$ and multiple system sizes given by the colour bar. (b)-(c): The same as (a) but instead for $\chi=0.5$ and $\chi=1$, respectively. In all panels, $\langle \mathrm{IPR}_4\rangle_{th}=1/2^{4}$, the expected thermal value of the $4$-qubit subsystem. Inset: Time average of the late-time \gls{ipr} dynamics shifted by $\langle \mathrm{IPR}_4\rangle_{th}$ as a function of system size taken between $t=30$ and $t=50$. All simulations were performed using ED with $B_z=1$ and we average over $100$ reference states and $10$ \gls{sk} graph instances.}
    \label{fig:IPRscaling}
\end{figure*}

For our cyclic algorithm to remain competitive for solving glassy problems, we must reliably cross the many-body localized to thermal phase transition when performing our Trotterised dynamics. Therefore, understanding the phase diagram of the \gls{sk} Hamiltonian in the presence of longitudinal and transverse fields becomes imperative for knowing whether our algorithm will be successful for a chosen set of parameters. However, determining the location of these critical points is notoriously difficult as it requires access to mid-spectrum eigenstates which can usually only be extracted using exact diagonalization (ED) techniques \cite{Pollman2014IsingMBL}. While tensor network methods like DMRG-X \cite{Khamenei2016DMRGX} have successfully been used to study mid-spectrum eigenstates in many-body localized Hamiltonians for large system sizes, tensor networks will notably struggle for spin-glass models due to the long range connectivity and random couplings in the graphs. Therefore, we propose a protocol to determine whether our chosen Hamiltonian parameters are located in the thermal phase, which can be directly implemented on quantum hardware.

To begin, for a given instance of an \gls{sk} graph, we choose our Hamiltonian parameters $B_z$, $\chi$, and reference state $S^{r}$. Our goal is to understand the phase diagram of the Hamiltonian reached at the end of step 2.\ of our protocol
\begin{equation}
\label{eq:PhasediagramHamiltonian}
    H=\sum_{1\leq i<j\leq N} J_{ij}Z_i Z_j -B_z\sum_{i=1}^N S^r_iZ_i -B_x\sum_{i=1}^N X_i,
\end{equation}
where $B_x=\chi B_z$, for graph sizes that are intractable using classical methods. A visualization of the phase diagram for small SK instances can be observed in fig.~\ref{fig:phase_diagram} with more details found in~\cite{SOM}. We note that the Hamiltonian in \cref{eq:PhasediagramHamiltonian} contains a continuous critical line separating the many-body localized and thermal regimes but for ease we solely focus on the case of $B_z=1$. Importantly, our algorithm hinges on whether we have successfully traversed deep into the thermal regime where the location of the critical point is observed to be $B_x \approx 0.25$~\cite{SOM}. 

To achieve this, we propose a short-time dynamics experiment and choose some thermalization metric to detect whether the dynamics have successfully thermalized. We choose to measure the \gls{ipr} on a subsystem $A$ \cite{Evers2000IPR,Evers2008IPR}, which is defined as
\begin{equation}\label{eq:IPRdef}
    \langle \mathrm{IPR}_{A}\rangle= \sum_{y \in A}p_{y}^2,
\end{equation}
where $p_y$ is the probability of measuring a given computational basis configuration of the subsystem, instead of other commonly used observables such as the spin imbalance. \gls{ipr} is also an efficient measure of delocalisation as the number of shots needed to resolve it with a fixed relative error scales with the size of the chosen subsystem~\cite{alam2026IPR}. As an example, for a marginal of size $|A|=4$ approximately $10^4$ shots are required to obtain a $\sim1\%$ relative error \cite{alam2026IPR}. Moreover, in accordance with predictions from Random Matrix Theory, the dynamics of chaotic Hamiltonians in the $t\rightarrow \infty$ limit thermalize to $\langle \mathrm{IPR}_{A}\rangle_{th}=1/2^{|A|}$ and provides a ``lower bound'' on thermalization which is generally reliable for local observables \cite{alam2026IPR}. Although our cyclic algorithm requires us to traverse the glassy landscape through multiple reference states, we show numerically that the choice of $S^{r}$ does not change the location of the critical point. See \cref{sec:level stats} of the SM~\cite{SOM} for details. Therefore, with this numerically informed rule of thumb, we only need to check the initial reference state, significantly reducing the quantum resources required for the entire cyclic algorithm.

\section{Results}\label{sec:results}

We first confirm our parameter testing algorithm discussed in \cref{subsec:parameter_scaling} by performing small scale dynamics simulations using exact diagonalization. We calculate the \gls{ipr} for 10 instances of \gls{sk} models between 6 and 16 nodes with $B_z=1$ and the system initialised in the product state $|\psi(0)\rangle=|01...01\rangle$. For each \gls{sk} instance, we average over 100 randomly chosen reference states and we take the $4$-qubit marginal. 

In \cref{fig:IPRscaling}(a) we plot the short-time dynamics of the scaled \gls{ipr} according to the expected thermal value $\langle \mathrm{IPR}_{A}\rangle_{th}$ for $\chi=0.25$, which is relatively close to $\chi_c \approx 0.25$ and also scale the dynamics by $N$. For these short-time scales, accessible on quantum hardware, we see a significant deviation from $\langle \mathrm{IPR}\rangle_{th}$, indicating that we are indeed quite close to the MBL-thermal critical line. Notably, in these short-time regimes we see that increasing the system size causes a much larger deviation, hinting that our cyclic algorithm could potentially fail for the chosen parameters.

In \cref{fig:IPRscaling}(b) we perform the same analysis for $\chi=0.5$ over the same \gls{sk} instances. Here, we observe excellent scaling in the number of nodes at later times, showing that the dynamics have successfully thermalized for all the system sizes considered. This indicates that, for \gls{sk} instances of sizes reachable using ED, $\chi=0.5$ would open a reasonably large gap that would hinder Landau-Zener transitions. We also observe similar behaviour for $\chi=1$, shown in \cref{fig:IPRscaling}(c), where all the dynamics have effectively thermalized at time $t=1$. This implies that $\chi=1$ would be an excellent parameter choice to perform our cyclic optimisation protocol, while also indicating that times which are experimentally accessible on quantum devices are sufficient for parameter testing.

In the inset of \cref{fig:IPRscaling}(c) we show the effect of performing long-time dynamics on the ability of the Hamiltonian to thermalize. In particular, we calculate $\langle \mathrm{IPR}_4\rangle_t$, the time averaged \gls{ipr} between times $t=30$ and $t=50$. Explicitly, we plot the absolute difference between $\langle \mathrm{IPR}_4\rangle_t$ and the expected thermal value $\langle \mathrm{IPR}_4\rangle_{th}$ as a function of system size for the same three values of $\chi$. For $\chi=0.25$ we observe a linear decrease towards $\langle \mathrm{IPR}_4\rangle_{th}$ whereas for $\chi=0.5$ and $\chi=1$ we observe excellent convergence to expected value even for small systems. While this demonstrates that long times are an excellent choice for studying the effects of thermalization, short-time dynamics are sufficient to determine the right Hamiltonian parameters for our protocol.

In addition to the MBL transition point, there are many other factors that affect the algorithmic performance. We perform extensive numerical investigations to address this in the SM~\cite{SOM}. More specifically, we detail the effects of Trotter order and step size in \cref{sec:Trotter order}, its dependence on the initial state in \cref{sec:reference state analysis}, and the ability to transition to lower energy eigenstates in \cref{sec:succes fss}. These are often performed with various values of $\chi$, which determines the evolution time for the final quench, and at various system sizes to elucidate finite size effects. In the next section, we use the understanding from these simulations to demonstrate that with an informed choice in the quench parameters, FRIO has the ability to successfully solve the max-cut problem on all-to-all connected while retaining non-adiabaticity.

\begin{figure}
    \centering
    \includegraphics[width=0.99\linewidth]{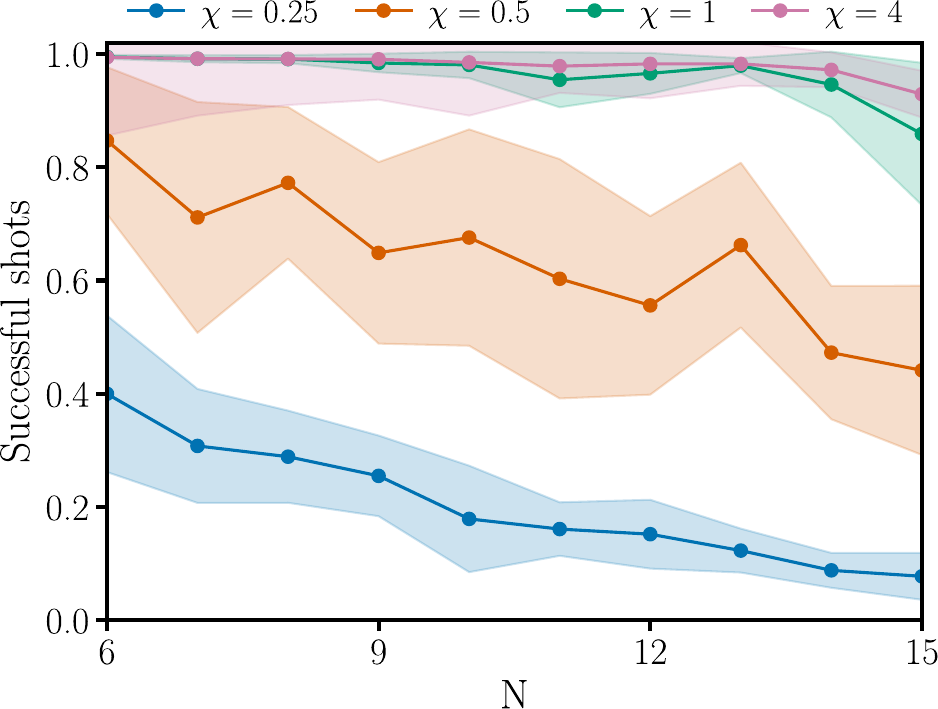}
    \caption{Fraction of the number of shots that successfully measure the ground state for an instances of \gls{sk} graphs using continuous FRIO. We choose multiple values of $\chi$ that lie close to the MBL transition point and in the thermal regime.  For each graph instance we randomly select $50$ starting reference states and allocate $1000$ shots to the FRIO protocol. All simulations have been performed for graphs with $N$ between $6$ and $15$ nodes and averaged over 10 graph instances, with the semi-transparent shaded areas representing the statistical error. We also set $B_z=1$.}
    \label{fig:Cont_resources}
\end{figure}

\subsection{Finding ground states using FRIO}

In previous sections we determined, via the \gls{ipr} metric, the location of the MBL phase transition for the 
\gls{sk} Hamiltonian given in \cref{eq:PhasediagramHamiltonian} for small graph instances which are classically simulable using ED. We also proposed a new method to perform efficient testing of potential parameters directly on quantum hardware, leveraging short-time dynamics and thermalization principles. We now demonstrate that with a careful choice of quench parameters, continuous time FRIO can be successfully leveraged to find the ground state of weighted graphs corresponding to the \gls{sk} Hamiltonian.

In \cref{fig:Cont_resources} we show the performance of FRIO using the continuous quench protocol defined in \cref{eq:annealing_path} for $4$ different choices of $\chi$. We choose the three values of $\chi$ which was included in our \gls{ipr} analysis and we add $\chi=4$, which is larger than $\chi_{\rm tri} \approx3.6$, the smallest $\chi$ required to encircle the system tricriticalpoint~\cite{Wang2022MBLopt}. We average over $10$ instances of the \gls{sk} model. For each instance we consider 50 randomly selected initial bitstrings $S^r$ and allocate $1000$ cycles for the FRIO protocol. As a performance metric we use the number of measurements that successfully measure the ground state energy which can be obtained using ED techniques. Here, we observe that $\chi=1$, the same value which was shown to be an excellent parameter choice via \gls{ipr} dynamics, consistently outperforms the two lower values of $\chi$ across the entire range of considered system sizes. Moreover, using $\chi=4$ offers no advantage over $\chi=1$, as it exhibits similar performance over the system sizes considered; we therefore restrict ourselves to $\chi=1$ for all further simulations.

In the next section, we use the understanding from these studies and compare our parameter-optimized digitised, cyclic annealing protocol against QAOA with optimal angles when solving the max-cut problem on all-to-all connected graphs described by the \gls{sk} model.


\begin{figure*}
    \centering
    \includegraphics[width=0.99\linewidth]{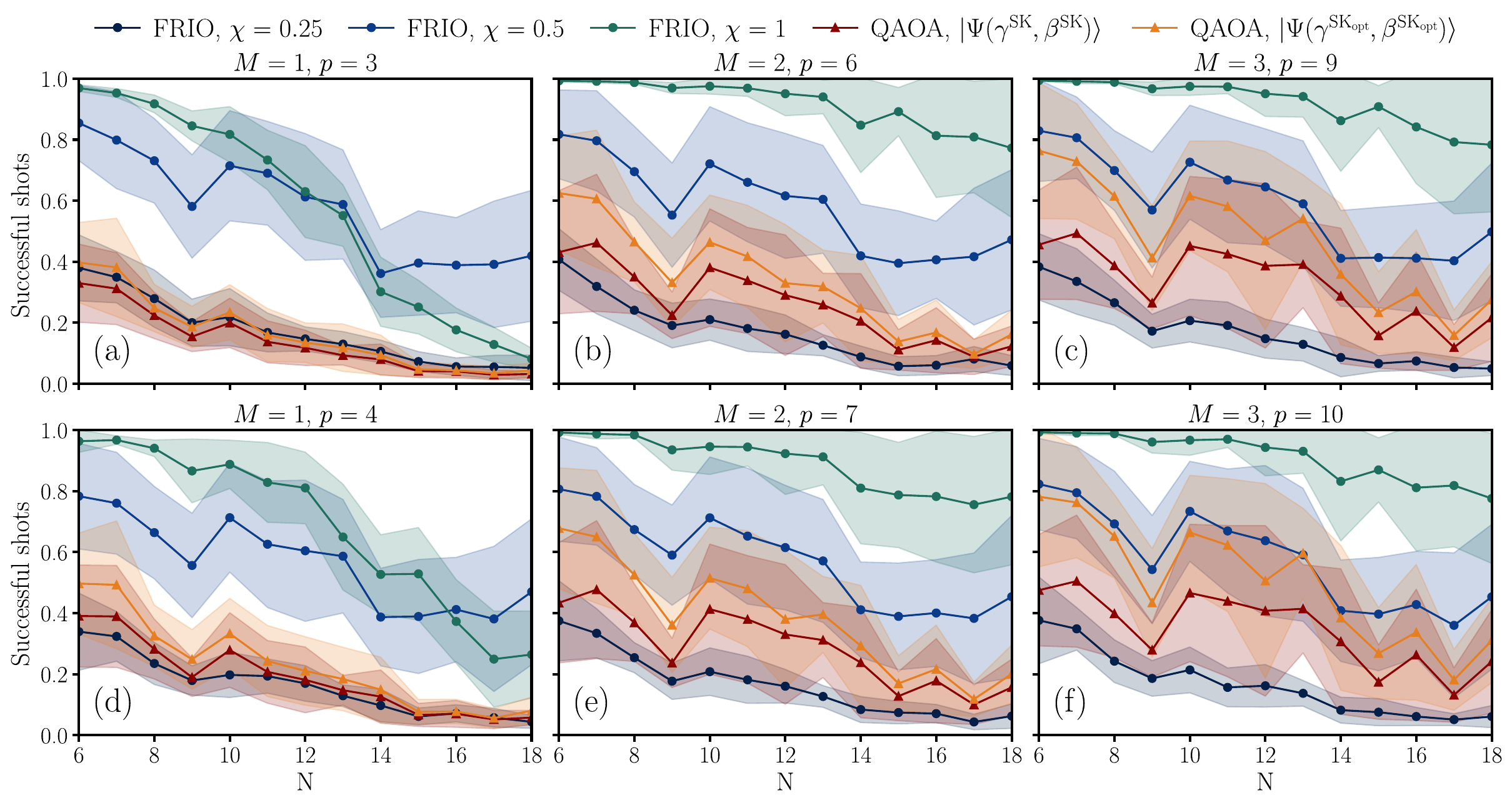}
    \caption{Fraction of the number of shots that successfully solve the max-cut problem for an instance of an \gls{sk} graph using digitised FRIO and a \gls{qaoa} circuit of the same depth of Trotter layers. Panels (a)-(c) use a first-order Trotter decomposition with $M=1,2,3$ Trotter steps for $\chi=0.25,0.5,1$ which is compared to a \gls{qaoa} circuit of depth $p=3,6,9$. Panels (d)-(f) use a second-order Trotter decomposition with $M=1,2,3$ Trotter steps and are compared to a \gls{qaoa} circuit of depth $p=4,7,10$. For the cyclic optimisation algorithm we randomly select 50 starting reference states and allocate 1000 shots to the cyclic optimisation protocol. For \gls{qaoa} we take $5\times10^4$ samples for each graph instance using the known angles from Refs.~\cite{Basso2022QAOAoptangles,Farhi2022quantumapproximate} and the angles after optimizing for each graph instances. All simulations have been performed for graphs with $N$ between 6 and 18 nodes and averaged over 10 graph instances, with the semi-transparent shaded areas representing the statistical error.}
    \label{fig:SK_shot_resources}
\end{figure*}

\subsection{Digitised cyclic optimisation vs QAOA}

We now demonstrate that, with a careful choice of quench parameters, digitised FRIO can directly out-perform \gls{qaoa} when solving the max-cut problem on graph instances corresponding to the quantum \gls{sk} model. For comparisons with \gls{qaoa}, we consider two choices of the \gls{qaoa} angles. The first are the angles $\gamma^{\text{SK}}_p$ and $\beta^{\text{SK}}_p$ introduced in~\cite{Basso2022QAOAoptangles,Farhi2022quantumapproximate}, which optimise the ensemble-averaged \gls{sk} model and were obtained via extrapolation methods. The second are the optimal angles $\gamma^{\text{SK}_\text{opt}}_p$ and $\beta^{\text{SK}_\text{opt}}_p$ for a single instance of the \gls{sk} model. To find these optimal angles we use $\gamma^{\text{SK}}_p$ and $\beta^{\text{SK}}_p$ as initial parameters in the optimisation landscape and optimise for the ground state energy. We emphasize that since all \gls{qaoa} angles are found classically we do not consider any quantum resources used during the optimisation procedure. However, at intractable system sizes, \gls{qaoa} will suffer an additional cost of parameter optimisation which is not present in digitised FRIO.

In \cref{fig:SK_shot_resources}(a) we compare  first-order Trotterised FRIO with $M=1$ Trotter step with $\chi=0.25,0.5,1$ against a \gls{qaoa} circuit of depth $p=3$ for multiple system sizes. As a comparison measure we use the number of shots that successfully sample the ground state energy as obtained using ED techniques. For FRIO we consider 50 randomly selected initial bitstrings $S^r$ as reference states with each $S^{r}$ allocated 1000 cycles. We perform a single measurement for each cycle and average over 10 instances of the \gls{sk} model, with the shaded areas in \cref{fig:SK_shot_resources} representing the statistical error in the performance from the averaging procedure. For each \gls{sk} instance we take $5\times10^4$ samples from the final \gls{qaoa} wavefunction and average over the same graph instances, thereby allocating the same number of shot resources for both FRIO and \gls{qaoa}. For all considered values of $\chi$, our algorithm either remains competitive or outperforms \gls{qaoa} for all considered graph sizes. Although for $\chi=0.25$ the system has not fully crossed the MBL-thermal critical line, we observe that cyclic optimisation is still competitive with \gls{qaoa}. Once $\chi$ has been increased deep into the thermal phase, our algorithm readily outperforms optimised \gls{qaoa} for all system sizes considered. For large systems, i.e., $N>13$, we see a large decrease in our algorithm's effectiveness which is due to the large Trotter error for such shallow circuits \cite{SOM}.

In \cref{fig:SK_shot_resources}(b)-(c) we perform the same analysis for $M=2$ and $M=3$ Trotter steps and compare to a \gls{qaoa} circuit of depth $p=6$ and $p=9$, respectively. Here, the value of $\chi$ has a greater effect on the number of successful shots: the smallest value, $\chi=0.25$ (i.e., the most non-adiabatic final quench), drives the success rate below that of conventional \gls{qaoa}, as one would expect. However, choosing a $\chi$ so that the system crosses the MBL to thermal phase transition almost guarantees that FRIO will beat \gls{qaoa}, even when the \gls{qaoa} circuits are individually optimized for a specific graph instance. Importantly, we find that for the quench durations chosen in this work, $\chi\approx1$, which is below the tricritical value $\chi_{\rm tri} \approx 3.6$ identified in Ref.~\cite{Wang2022MBLopt}, is sufficient to ensure a high success probability across all system sizes considered. This suggests that our protocol does not rely on circling the tricritical point between glassy, MBL, and thermal phases, as required in Ref.~\cite{Wang2022MBLopt}. Moreover, as also discussed in \cref{sec:succes fss}, FRIO is capable of lowering the energy of the system even at very low energy density. This indicates that the theory we developed in \cref{subsec:nonadiabatic} gives a useful physical picture of the mechanism underpinning our algorithm, but does not fully capture its behaviour. Importantly, for the chosen quenches, which are far from adiabatic, the Trotterisation remains accurate enough despite the small number of steps used. In particular, we highlight that unlike Ref.~\cite{Wang2022MBLopt}, we achieve good convergence to the ground state using exclusively non-adiabatic quenches. One possible explanation for the resilience to Trotterisation is that the latter could introduce counterdiabatic terms which would improve the performance for low-depth circuits~\cite{Wurtz2022}. We leave the error analysis and potential origin of counterdiabatic terms to a future investigation.

In \cref{fig:SK_shot_resources}(d)-(f) we perform the same analysis for the second-order Trotterised algorithm $\mathrm{FRIO}_2$ with $M=1$, $M=2$, and $M=3$ Trotter steps,  respectively, and compare to \gls{qaoa} circuits of depth $p=4$, $p=7$, and $p=10$. Overall we find that using a second-order Trotter decomposition adds no notable benefit over using a first-order one, except when we consider graphs with $N\geq16$ nodes, where $\mathrm{FRIO}_2$ shows a better performance over standard QAOA. This suggests that the dynamics for both the first- and second-order decompositions are relatively similar, especially for $M=2$ and $M=3$ Trotter steps; see \cref{sec:Trotter order} in the SM for more details \cite{SOM}. Therefore, it may be beneficial to allocate more resources to increase the number of Trotter steps for a first-order decomposition or the ability to quench further into the thermal regime. This could, in turn, allow our algorithm to converge to the ground state faster and with fewer shot resources, but we leave this to future investigation. 

\section{Conclusions}\label{sec:conclusion}

In this work, we introduce a digitised iterative non-adiabatic cyclic algorithm for finding ground states in all-to-all connected graphs described by the quantum \gls{sk} model. By performing a sequence of dynamical quenches that traverse the phase diagram across the MBL-thermal transition and back, we demonstrate excellent convergence to the ground state for graph sizes that are tractable using classical simulation techniques. When considering \gls{qaoa} circuits of a similar depth, our algorithm outperforms optimized \gls{qaoa} in the number of useful shot resources if appropriate quench parameters are correctly identified. Moreover, we also introduced a method for finding these quench parameters based on short-time thermalization dynamics. This has the unique advantage of being easily implementable on quantum hardware compared to traditional thermalization measures with favourable resolution scaling to systems unreachable using classical techniques.

While the advantage of our algorithm over \gls{qaoa} has been demonstrated for classically tractable \gls{sk} instances, the parameters chosen throughout this work may not carry over to larger graphs, particularly for low-depth circuits. For larger systems, we speculate that longer quench durations may become necessary, requiring more Trotter steps to keep the Trotter error under control. A more rigorous analysis of the relationship between the quench duration and the Trotter decomposition would therefore be valuable, and would also improve our understanding of the potential counterdiabatic effect in our protocol. 

In principle, our algorithm could be implemented on current quantum annealing platforms but will be limited to certain quench sequences. One of the main advantages of digitisation is the freedom to choose more complex trajectories, add extra tuning parameters, or even blend different algorithms within the protocol, such as imaginary time evolution, which cannot be as easily implemented on quantum annealers. If Trotterisation does introduce a counterdiabatic effect, this would allow the use of highly non-adiabatic quenches, allowing digital quantum computers to become an excellent tool for studying faster, non-adiabatic annealing protocols. Finally, since our algorithm does not require variational optimisation, it is well suited for direct implementation for solving optimisation problems on platforms with a limited shot budget, such as trapped ion or neutral atom processors, where their reconfigurable all-to-all connectivity can also be leveraged.

\section{AI use}

\Cref{subsec:nonadiabatic} was partially developed through discussions with Claude. Claude was also used to identify typographical errors throughout the manuscript. All AI-assisted outputs were critically reviewed by the authors, who take full responsibility  for the accuracy, integrity and final content of the work.

\section{Acknowledgments}

We would like to thank Ieva \v{C}epait\.{e}, Elizabeth Crosson, Niam Vaishnav and Andrew Hallam for helpful discussions. We also thank Leo Zhou for sharing the optimal angles and best practices for QAOA.  The cyclic optimisation algorithm was implemented in Qiskit~\cite{qiskit2024} and the exact diagonalization simulations were performed in QuSpin~\cite{QuSpin1}. R.S. acknowledges support by the Leverhulme Trust Research Leadership Award RL-2019-015 and EPSRC Grants No. EP/Z533634/1 and No. UKRI1337.

\bibliography{references}

\newpage 
\cleardoublepage

\onecolumngrid

\begin{center}
\textbf{\large Supplemental Material for ``Non-adiabatic Optimisation via Localisation–Delocalisation Transitions"}\\[5pt]
\vspace{0.1cm}

\end{center}

\setcounter{equation}{0}
\setcounter{figure}{0}
\setcounter{table}{0}
\setcounter{page}{1}
\setcounter{section}{0}

\makeatletter
\let\addcontentsline\latex@addcontentsline
\makeatother

\renewcommand{\theequation}{S\arabic{equation}}
\renewcommand{\thefigure}{S\arabic{figure}}
\renewcommand{\thesection}{S\arabic{section}}
\renewcommand{\thepage}{\arabic{page}}
\renewcommand{\thetable}{S\arabic{table}}

\tableofcontents

\section{Derivation of the master equation}\label{app:Master_eq}

Let $|\psi_0\rangle$ be the initial (reference) state, with classical
energy $E_r:=\langle\psi_0|H_{\rm SK}|\psi_0\rangle$, the evolved state expanded in the bitstring basis $|b_i\rangle$ is
\begin{align}
    |\Psi(t)\rangle&=U(t)|\psi_{0}\rangle=\sum_{r}\langle b_{r}|U(t)|\psi_{0}\rangle|b_{r}\rangle=\sum_{r}c_{r}(t)|b_{r}\rangle.
\end{align}
The success probability is given by
\begin{align}
    P(t)&=\sum_{n:E_{n}(t) < E_{r}}|\langle b_{n}|U(t)|\psi_{0}\rangle|^{2}=\sum_{n:E_{n}(t)< E_{r}}c_{n}^{*}(t)c_{n}(t)
\end{align}
and satisfies the differential equation
\begin{align}
    \frac{dP}{dt}&=\sum_{n:E_{n}(t)< E_{r}}c_{n}^{*}(t)\frac{d}{dt}c_{n}(t)+\frac{d}{dt}c_{n}^{*}(t)c_{n}(t)\\\frac{d}{dt}c_{n}(t)&=\frac{d}{dt}\langle b_{n}|U(t)|\psi_{0}\rangle=-i\langle b_{n}|H(t)U(t)|\psi_{0}\rangle=\sum_{m}-i\langle b_{n}|H(t)|b_{m}\rangle\langle b_{m}|U(t)|\psi_{0}\rangle\\
    &=\sum_{m}M_{nm}(t)c_{m}(t),
\end{align}
where we have defined $M_{nm}:=i\langle b_{n}|H(t)|b_{m}\rangle$. A straightforward manipulation using that $M_{ij}^*=-M_{ji}$ leads us to
\begin{align}\label{eq:dpdt}
    \frac{dP}{dt}=\sum_{\substack{n:E_{n}(t)< E_{r}\\
m:E_{m}(t)\geq E_{r}}}\left(M_{nm}(t)c_{n}^{*}(t)c_{m}(t)-M_{mn}(t)c_{m}^{*}(t)c_{n}(t)\right).
\end{align}
\Cref{eq:dpdt} has a natural interpretation of a continuity equation for the probability, that can only change if there is a current (described by $M_{nm}$) driving energy levels across the energy boundary $E_r$. As the system size $N$ increases and the number of energy levels increases exponentially with $N$ we look at a coarse-grained description of this probability. To do this we introduce the density of states 
 \begin{align}
     \mathcal{P}(E,t)=\sum_n |c_n(t)|^2\delta(E-E_n(t)),
 \end{align}
 with $\int_{-\infty}^{E_r}dE\mathcal{P}(E,t)=p(t)$, and model the change in probability through the master equation
 \begin{align}\label{eq:ME}
     \partial_t\mathcal{P}(E,t)=\int dr \left[W(E-r,r)\mathcal{P}(E-r,t) - W(E,r)\mathcal{P}(E,t)\right],
 \end{align}
 where $W(E,r)$ is the jump rate associated with the energy jump from $E$ to $E+r$. Assuming a smooth dependence of the jump rates in the energy, we expand in a Taylor series 
 \begin{align}
     W(E-r,r)\mathcal{P}(E-r,t)=\sum_{k=0}^\infty\frac{(-1)^kr^k}{k!}\partial_E^k(W(E,r)\mathcal{P}(E,t)),
 \end{align}
obtaining the Kramers-Moyal expansion \cite{Kramers_1940,Moyal_1949}
\begin{align}
    \partial_t\mathcal{P}(E,t)=\sum_{k=1}^\infty\frac{(-1)^k}{k!}\partial_E^k\left(A^{(k)}(E)\mathcal{P}(E,t)\right), \quad\mbox{with}\quad A^{(k)}(E)=\int dr {r^k}W(E,r).
\end{align}
 Keeping the first two terms in the expansion produces the Fokker-Planck equation
 \begin{align}\label{eq:FP}
     \partial_t \mathcal{P}(E,t) = -\partial_E\big[A^{(1)}\mathcal{P}\big] + \frac{1}{2}\,\partial_E^2\big[A^{(2)}\mathcal{P}\big],
 \end{align}
 which we use to model the evolution of the probability of observing a state of energy $E$ at time $t$. Truncating at second order allows us to solve this equation given the diffusion and drift constants and is motivated by the assumption of many levels being crossed during the sweep. Under this assumption, an Edgeworth series \cite{Edgeworth1905,Cramer1946} can be defined in terms of a coarse-grained energy variable that corresponds to the sum of $n_c$ (assumed) identical and independent random energy contributions. By the central limit theorem the distribution associated with this coarse-grained energy approaches a Gaussian distribution with corrections that decay as $\mathcal{O}(1/\sqrt{n_c})$, whose behaviour is governed by the above Fokker-Planck equation.

 \subsection{Modeling the diffusion as Poisson process}
 \label{app:rates_Poisson}

 The Fokker-Planck \cref{eq:FP} depends on the functions $A^{(1)}$ and $A^{(2)}$ which parameterize the drift and diffusion of the underlying stochastic process respectively. We model the stochastic process of energy jumps across the ramp as a Poissonian process \cite{DaleyVereJones2003}. Let us consider the adiabatic basis, which is the basis associated with the instantaneous eigenvalues. As we sweep $B_z$ and $B_x$ back to the original SK Hamiltonian, at an avoided level crossing we can have one of the following cases
 \begin{enumerate}
     \item The system remains in the original state (with probability $1 -\langle p_{\rm nLZ}\rangle $),
     \item The system jumps with probability proportional to $\langle p_{\rm nLZ}\rangle$. 
Whenever the system jumps it can either gain energy if it is going from a lower to a larger instantaneous eigenstate, or it can lose energy if the jump is from a high energy state to a low energy one.
 \end{enumerate}
Modeling this as a Poisson process is a consequence of the following theorem, which can be thought of as a central limit theorem for point processes

\begin{theorem}[Superposition theorem for point processes; \cite{Grigelionis1963,DaleyVereJones2003,Kingman1993}]
\label{thm:superposition}
For each $n$, let $N_1^{(n)},\dots,N_n^{(n)}$ be independent point processes
on $\mathbb{R}$, and set $N^{(n)} = \sum_{i=1}^n N_i^{(n)}$. Suppose that for
every bounded Borel set $B$:
\begin{enumerate}[i)]
  \item \textbf{Uniform asymptotic negligibility:}
  $\lim_{n\rightarrow \infty}\sup_{1\le i\le n} P\big(N_i^{(n)}(B)\ge1\big)=0$;
  \item \textbf{Convergent total intensity:}
  $\lim_{n\rightarrow \infty}\sum_{i=1}^n P\big(N_i^{(n)}(B)\ge1\big)=\rho(B)$
  for some Radon measure $\rho$;
  \item \textbf{Negligible multiplicities:}
  $\lim_{n\rightarrow \infty}\sum_{i=1}^n P\big(N_i^{(n)}(B)\ge2\big)=0$.
\end{enumerate}
Then $N^{(n)}$ converges in distribution (in the sense of convergence of
finite-dimensional distributions, equivalently weak convergence on the space
of counting measures) to a Poisson process with intensity measure $\rho$.
\end{theorem}
 
\noindent
Condition (iii) is automatic in our setting: a given competing local
minimum and the reference curve
$\tilde E_r$ are each locally monotonic, so they cross at most once in
a bounded window. Conditions
(i)--(ii) become \emph{more} accurate as $N$ grows: for (i) each individual local
minimum's energy is an essentially independent draw from the SK complexity
spectrum, so the chance
that any one specific minimum crosses within a given window shrinks. For (ii) the total count of minima
$\mathcal N(N)\sim e^{N\,s(\epsilon)}$ grows, keeping their sum
$\rho(N)=\sum_l P(\text{minimum }l\text{ crosses})$ a well-defined quantity. 

 By the \textbf{Poisson thinning (splitting) theorem} \cite{Kingman1993}, independently
classifying the points of a Poisson process into categories with fixed
probabilities produces \emph{independent} Poisson sub-processes, one per
category, each with rate equal to the original rate times that category's
probability:
\begin{equation}
  \rho_{\rm none} = \rho\,(1-\langle p_{\rm nLZ}\rangle), \qquad
  \rho_{\downarrow} = \rho\,\langle p_{\rm nLZ}\rangle\frac{N(E<E_r)}{N(E<E_r)+N(E\geq E_r)}, \qquad
  \rho_{\uparrow} = \rho\,\langle p_{\rm nLZ}\rangle\,\frac{N(E\geq E_r)}{N(E<E_r)+N(E\geq E_r)}.
\end{equation}
In the last two rates, we have introduced the fraction of states below and above the reference energy to qualify the processes going down or up in energy respectively.

It is good to clarify why $\langle p_{\rm nLZ} \rangle$ is the correct probability to consider. Focusing on a single crossing between two levels $|A\rangle,|B\rangle$, as in the usual frame of Landau-Zener transitions, the Hamiltonian has the form
\begin{align}
    H_2=\left(\begin{matrix}
     vt & \Delta \\
     \Delta & -vt
     \end{matrix}\right)
\end{align}
written in the diabatic basis $|A\rangle$ and $|B\rangle$. Now let us imagine that both $|A\rangle$ and $|B\rangle$ are dominated by a single bitstring configuration. At the asymptotic limit $t\rightarrow \pm \infty$ (corresponding to the beginning and end of the annealing protocol), the eigenstates are the diabatic basis. Note that an {\it adiabatic} transition changes the diabatic labels, so an adiabatic transition with probability $\langle p_{\rm nLZ}\rangle$ induces a hop between bitstrings, which are the basis states where the energy of the $H_{\rm SK}$ Hamiltonian is measured at the end.

Finally we can convert these rates into the average energy and variance of the process using that at each crossing the system might jump and gain or lose energy, so we have
\begin{align}
  \langle \Delta E\rangle &= \bar\Delta(\rho_\downarrow - \rho_\uparrow)t_f
  = \rho\langle p_{\rm nLZ}\rangle\bar\Delta\eta t_f := A^{(1)}t_f, \\
  \mathrm{Var}(\Delta E) &= \bar\Delta^2(\rho_\downarrow + \rho_\uparrow)t_f
  = \rho\langle p_{\rm nLZ}\rangle\,\bar\Delta^2 t_f := A^{(2)} t_f.
\end{align}

\subsection{Assumptions behind the Gaussian approximation}
\label{sec:validity}

The truncation at second order in the Kramers-Moyal expansion, and the resulting
Gaussian propagator, require:
\begin{enumerate}
  \item \textbf{Many crossings within $\tau_3$}: $\rho\,\tau_3\gg1$, so that the
  central-limit argument underlying the Gaussian
  approximation applies. 
  \item \textbf{Approximate independence of successive crossings}: correlations
  between nearby gaps (beyond the local level-repulsion captured in the Wigner
  surmise at a single crossing) are neglected.
  \item \textbf{Slowly varying $A^{(1)},A^{(2)}$ over the sweep}: justified in a
  window close to $\chi_{\rm tri}$ where the local statistics change smoothly, but not
  necessarily far from it.
\end{enumerate}

\subsection{Low energy minima hybridisation}\label{sec:hybrid}

In \cref{subsec:nonadiabatic} of the main text, we demonstrated that during the performance of a single cycle of FRIO, we achieve a non-zero success probability of transitioning to lower energy states which are then used for the next optimisation cycle. However, FRIO is expected to slow down and could potentially become stuck in low-lying energy minima that do not correspond to the ground state. We now show that it is still possible to continually transition to even lower energy states by repeating the analysis in \cref{fig:spectral} for reference states that correspond to extremely low-lying excited eigenstates.

\begin{figure}
    \centering
    \includegraphics[width=0.99\linewidth]{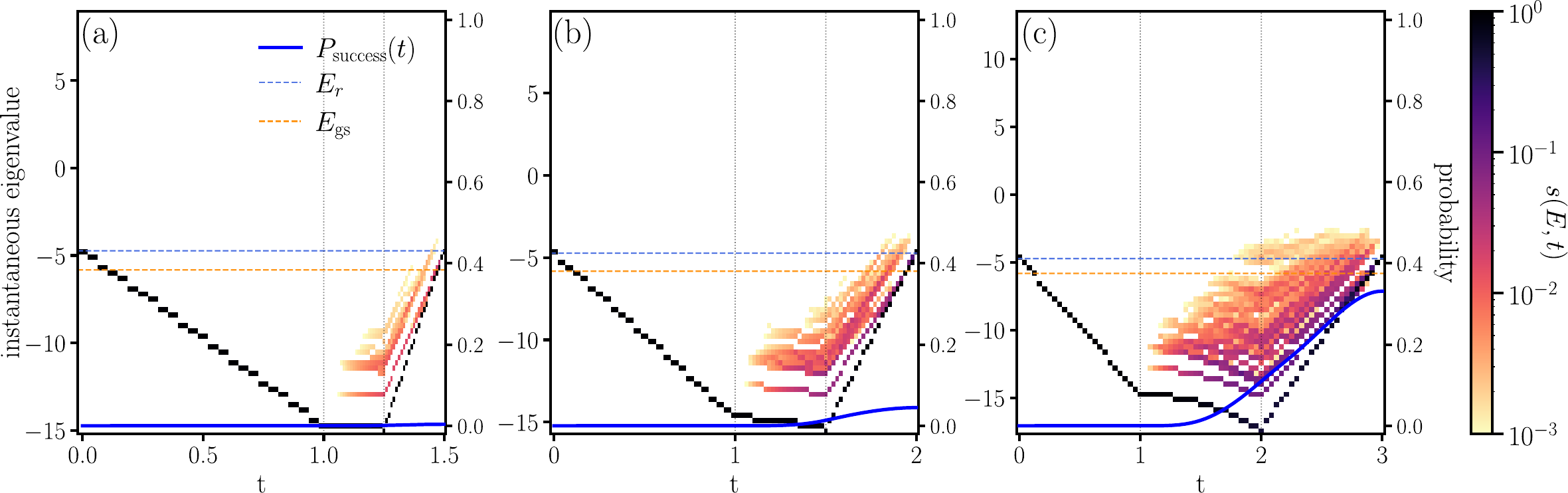}
    \caption{Evolution of the spectral weight $s(E,t)=\sum_n|\langle n(t)|S^r\rangle|^2\delta(E-E_n(t))$ in the FRIO algorithm for a reference state that corresponds to the second excited eigenstate of the \gls{sk} instance. Panels (a)-(c) correspond to a single optimisation cycle with $\chi=0.25$, $\chi=0.5$, and $\chi=1$, respectively. Here $|n(t)\rangle$ is the instantaneous eigenvector with instantaneous eigenvalue $E_n(t)$. In blue we plot the probability of a successful energy transition, $P_{\rm success}$, during the ramping sequence associated with each value of $\chi$. The dashed blue line marks the energy of the initial state, while the orange dashed line marks the ground-state of $H_{\rm SK}$. The vertical grey dashed lines mark the beginning of the second and third step of the algorithm.}
    \label{fig:low energy minima}
\end{figure}

In \cref{fig:low energy minima} we plot the evolution of the spectral weight during one cycle of FRIO where the initial reference state corresponds to the second excited eigenstate of the considered \gls{sk} instance of size $N=10$. Panels (a)-(c) correspond the quench parameters $\chi=0.25$, $\chi=0.5$, and $\chi=1$, respectively, while we set $B_z=1$. Just like we observed in \cref{fig:spectral} in the main text, we see that the first quench pushes the initial state to the bottom of the spectrum while the second quench causes the reference state to hybridize with nearby energy minima. The final quench, which returns to the original Hamiltonian, is required to ensure that we measure a new bit string with lower energy. Similarly, from the success probability, we can see that with the correct choice of quench parameters, even reference states corresponding to low energy eigenstates can still successfully transition to lower energy minima, albeit at lower rate. While this analysis is confined to a single \gls{sk} instance with a fixed number of nodes, in \cref{sec:succes fss} we show that this behaviour extends to larger \gls{sk} graphs using finite-size scaling techniques.

\section{A second-order Trotterisation of FRIO}\label{sec:Trotter 2}

In the main text we considered a digitised version of FRIO suitable for applications on digital quantum hardware via a first-order Trotterisation. However, there are many possible decompositions which will give a better performance at the expense of requiring more Trotter layers for execution. Therefore, we expand the analysis performed in \cref{subsec:Trotter 1} to second order.

To begin, we repeat the unitary evolution which describes the triangular path which traverses the phase diagram. The continuous time evolution unitary is 

\begin{align}
    U_\triangle(t)=\begin{cases}
        U_1(t) \quad &\mbox{for $t\in[0,t_1]$},\\
        U_2(t_2-t_1)U_1(t_1) \quad &\mbox{for $t\in[t_1,t_2]$},\\
        U_3(t-t_3)U_2(t_2-t_1)U_1(t_1)\quad  &\mbox{for $t\in[t_2, t_3]$},
    \end{cases}
\end{align}

where $U_i(t)=\mathcal{T}e^{-i\int_0^{t}H_i(t')dt'}$ is the time ordered evolution. As we did in the main text, we approximate each $U_{i}(t)$ using a second-order Trotterisation. Therefore, the three evolution unitaries are 

\begin{align}\label{eq:second_order_Trotter_decomp}
    \centering 
        U^{\rm Tr}_1(t)&=e^{-iH_{\rm SK} t} e^{-iH_z \int_{0}^{t}B_{z}(t')dt'},\\
        U^{\rm Tr}_2(t)&=\prod_{k=1}^{M}(e^{-i\frac{t}{2M}H_{x}B_x(\frac{kt}{M} +\frac{kt}{2M})} e^{-iH_{\rm SK}\frac{t}{M}} e^{-iB_zH_z\frac{t}{M}} e^{-i\frac{t}{2M}H_{x}B_x(\frac{kt}{M} +\frac{kt}{2M})}),\\
        U^{\rm Tr}_3(t)&=\prod_{k=1}^{M}(e^{-i\frac{t}{2M}H_{x}B^{(3)}_x(\frac{kt}{M} +\frac{kt}{2M})} e^{-iH_{\rm SK}\frac{t}{M}} e^{-iB^{(3)}_z(\frac{kt}{M} +\frac{kt}{2M})H_z\frac{t}{M}} e^{-i\frac{t}{2M}H_{x}B^{(3)}_x(\frac{kt}{M} +\frac{kt}{2M})}),
\end{align}
which uses $M$ layers of the evolution for $U_{2}(t)$ and $U_{3}(t)$. For ease we will call this $\mathrm{FRIO}_2$.

We can compare the results of $\mathrm{FRIO}_2$ with a \gls{qaoa} circuit of depth $p=3M+1$ where $p$ is the \gls{qaoa} depth. For example, $\mathrm{FRIO}_2$ with $M=1$ Trotter step is directly comparable to a $p=4$ \gls{qaoa} circuit. Moreover, just like our standard FRIO algorithm, $\mathrm{FRIO}_2$ also obeys an $\mathcal{O}(N^2)$ scaling in the number of two-qubit gates required to execute the circuit on hardware with all-to-all connectivity.

\section{Weighted $k-$regular graphs as spin glass problems}

In the main text we focused on solving the max-cut problem for all-to-all connected graphs that are described by the \gls{sk} model. In this section, we will demonstrate that with a careful choice of parameters, FRIO can also out-perform \gls{qaoa} circuits of a similar depth when finding ground states of a variety of spin glass models. We also show that by considering short-time dynamics experiments and local thermalization observables which can be implemented on current quantum hardware, we can accurately test that our chosen quench parameters are located in the thermal phase. 

We introduce a variant of the \gls{sk} model where we limit the graph to only have $k$ connections. The weighted $k$-regular Hamiltonian is defined similarly to the SK model,

\begin{equation} \label{eq:klocHam}
    H_{k}=\sum_{1\leq i<j\leq N}A^{(k)}_{ij}J_{ij}Z_i Z_j,
\end{equation}
where $A^{(k)}_{ij}$ is the adjacency matrix describing the layout of the connections between the nodes and $k$ is the number of edges from each node. Similarly to the \gls{sk} model, $J_{ij}$ is the cost matrix whose elements are sampled from a normal distribution with zero mean and variance $J^2/N$, while all $2^N$ eigenstates are also encoded by bitstrings. Moreover, since the elements of the cost matrix are also randomly selected, the energies also form a glassy energy landscape which would be perfectly suited for our digitised cyclic algorithm. Since our study is limited in the number of nodes that we can exactly simulate, we choose to restrict ourselves to only $k=3,4,5$ regular graphs in which connections between nodes are also generated randomly. We note that for odd $k$, we can only consider graphs with an even number of nodes.

Alongside the SK model, the \gls{qaoa} angles which on average best optimize eq.~\eqref{eq:klocHam} are also already known for low depth \gls{qaoa} circuits. We use the \emph{\bf{SKatan}} method ~\cite{Shaydulin2023skatan,Sureshbabu2024skatan,cepaite2025QAOAwarmstarts} which relies on the fact that in the $k\rightarrow\infty$ limit a weighted $k$-regular graph approaches the \gls{sk} model. Since $\gamma^{\text{SK}}$ and $\beta^{\text{SK}}$ are the angles which on average best optimize the \gls{sk} Hamiltonian as used in the main text, we can easily reweight them to find the angles that on average best optimize eq.~\eqref{eq:klocHam} as:

\begin{equation}
    (\gamma^{k},\beta^{k}) = (\gamma^{\text{SK}}\arctan(\frac{1}{\sqrt{k-1}}),\beta^{\text{SK}}),
\end{equation}
which will have the final state $\lvert \Psi(\gamma^{k},\beta^{k})\rangle$.
We also consider the approach where the \gls{qaoa} angles are classically optimized for each instance of eq.~\eqref{eq:klocHam} using $\gamma^{k}$ and $\beta^{k}$ as a starting point and have the corresponding final state $\lvert \Psi(\gamma^{k_\text{opt}},\beta^{k_\text{opt}})\rangle$. Like the SK model, since we optimize $(\gamma^{k_\text{opt}},\beta^{k_\text{opt}})$ classically, there are no quantum resources spent during the optimisation procedure. Finally, we note that we could choose other parameter setting techniques such as reweighting the optimal angles of the unweighted maxcut-problem \cite{Shaydulin2023skatan} and the \emph{\bf{Dweight}} method which is built from the fixed angle conjecture \cite{Wurtz2021Fixedangle}.

\begin{figure*}
    \centering
    \includegraphics[width=0.9\linewidth]{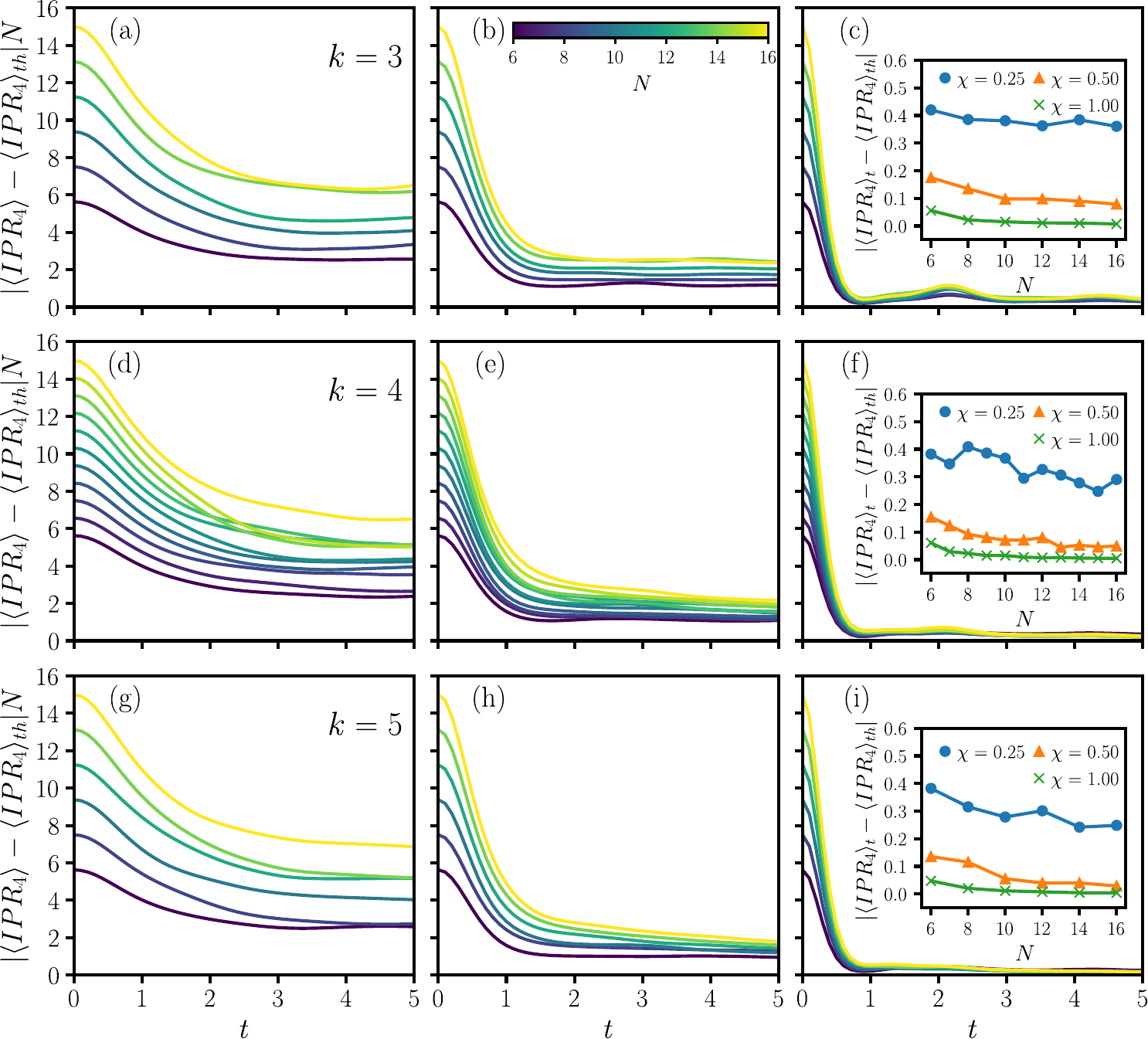}
    \caption{Marginal \gls{ipr} dynamics for parameter testing experiments using weighted $k$-regular type graphs. (a-c): The short-time dynamics of the rescaled \gls{ipr} of $k=3$ regular graphs for a $4$-qubit marginal for $\chi=0.25$, $\chi=0.5$, and $\chi=1$ respectively. Multiple system sizes are plotted given by the colour bar in panel (b). (d-f): The same as panels (a-c) but for weighted $k=4$ regular graphs. (g-i): Also the same as panels (a-c) but for weighted $k=5$ regular graphs. Here, $\langle \mathrm{IPR}_4\rangle_{th}=1/2^{4}$, the expected thermal value of the $4$ qubit subsystem. Insets: Time average of the late-time \gls{ipr} dynamics shifted by $\langle \mathrm{IPR}_4\rangle_{th}$ as a function of system size taken between $t=30$ and $t=50$. The panel the inset is located in corresponds to the $k$-regular graph simulation of that panel. All simulations were performed using ED with $B_z=1$ and we average over $100$ reference states and $10$ weighted $k$-regular graph instances.}
    \label{fig:IPRscaling_dreg}
\end{figure*}

\subsection{Parameter testing using quantum thermalization}
Just like we did in the main text, for a given instance of a weighted $k$-regular graph, we choose our Hamiltonian parameters $B_z$, $\chi$, and reference state $S^{r}$. Therefore, we want to understand the phase diagram of the weighted $k$-regular Hamiltonian
\begin{equation}
\label{eq:PDHamiltonianklocal}
    H=\sum_{1\leq i<j\leq N} A^{(k)}_{ij}J_{ij}Z_i Z_j -B_z\sum_{i=1}^N S^r_iZ_i -B_x\sum_{i=1}^N X_i,
\end{equation}
where $B_x=\chi B_z$ and we set $B_z=1$. To test whether Hamiltonians described by \cref{eq:PDHamiltonianklocal} contain a localized to thermal phase transition, we perform a similar analysis to fig.~\ref{fig:IPRscaling} using our parameter selection algorithm based on the \gls{ipr} of a subsystem $A$ defined by \cref{eq:IPRdef} (see Sec.~\ref{subsec:parameter_scaling} in the main text for further details). For classically simulable instances of weighted $k$-regular graphs we perform small scale dynamics simulations of the \gls{ipr} using exact diagonalization. We randomly generate 10 instances of \cref{eq:klocHam} between $N=6$ and $N=16$ nodes and average over 100 randomly chosen reference states for each graph instance. The system is initialised in the product state $|\psi(0)\rangle=|01...01\rangle$ and we calculate the IPR of the $4$-qubit marginal.

In \cref{fig:IPRscaling_dreg}(a) we plot the short-time dynamics of the scaled \gls{ipr} for $k=3$ regular Hamiltonians at $\chi=0.25$. For $\chi=0.25$, the short-time scales that are accessible on quantum hardware show a dramatic deviation from $\langle \mathrm{IPR}_4\rangle_{th}$, indicating that we are quite close to the MBL-thermal critical line. Just like for the SK model, we observe that increasing the system size causes a larger deviation, which suggests that the chosen parameters could fail. In \cref{fig:IPRscaling_dreg}(b) we perform the same analysis for $\chi=0.5$ which shows a smaller deviation from $\langle \mathrm{IPR}_4\rangle_{th}$, indicating that $\chi=0.5$ lies much further away from the critical line and would potentially be a better choice of parameters. However, the best scaling is in \cref{fig:IPRscaling_dreg}(c) which is for $\chi=1$, showing that the dynamics of the \gls{ipr} have successfully converged to $\langle \mathrm{IPR}_4\rangle_{th}$ during these short time scales. In the inset of \cref{fig:IPRscaling_dreg} we calculate $\langle \mathrm{IPR}_4\rangle_{t}$, the time-averaged \gls{ipr} between times $t=30$ and $t=50$, as a function of system size for the same three values of $\chi$. For $\chi=0.25$ we observe an extremely slow linear decrease towards $\langle \mathrm{IPR}_4\rangle_{th}$ whereas for $\chi=0.5$ there is a faster linear decrease towards $\langle \mathrm{IPR}_4\rangle_{th}$. Finally for $\chi=1$ we see excellent convergence to the expected value at small systems.

In \cref{fig:IPRscaling_dreg}(d-f) we perform the same analysis as above but for $k=4$ regular Hamiltonians in which we observe a similar pattern of behaviour in both the short- and long-time dynamics of the \gls{ipr}. We show that by increasing $\chi$ we see a significant decrease in the deviation from the expected thermal value of the \gls{ipr}, with the dynamics having converged by $\chi=1$. This is further demonstrated in the inset of \cref{fig:IPRscaling_dreg}(f) which shows that the long-time dynamics demonstrate excellent finite-size scaling of $\langle \mathrm{IPR}_4\rangle_{t}$, with $\chi=0.25$ showing a linear decrease towards the expected value whereas the other two values of $\chi$ have converged for large enough system sizes. Finally, in \cref{fig:IPRscaling_dreg}(g-i) we perform the same analysis for $k=5$ regular Hamiltonians obtaining similar behaviour in both the short-time and the long-time dynamics of the \gls{ipr} with $\chi=1$ always being the best choice in the parameters to make as it gives us the best convergence on the time scales accessible with quantum hardware. We highlight that the efficacy of our parameter testing algorithm becomes significantly better as we increase the number of graph edges $k$, indicating that the speed at which these types of Hamiltonians thermalize is dependent on the connectivity of the graph. While we only perform a simple analysis, it would be beneficial to accurately study the thermalizing behaviours of $k$-regular graph Hamiltonians, which we leave for future work.  

\subsection{Outperforming QAOA for glassy problems.}

\begin{figure*}
    \centering
    \includegraphics[width=0.99\linewidth]{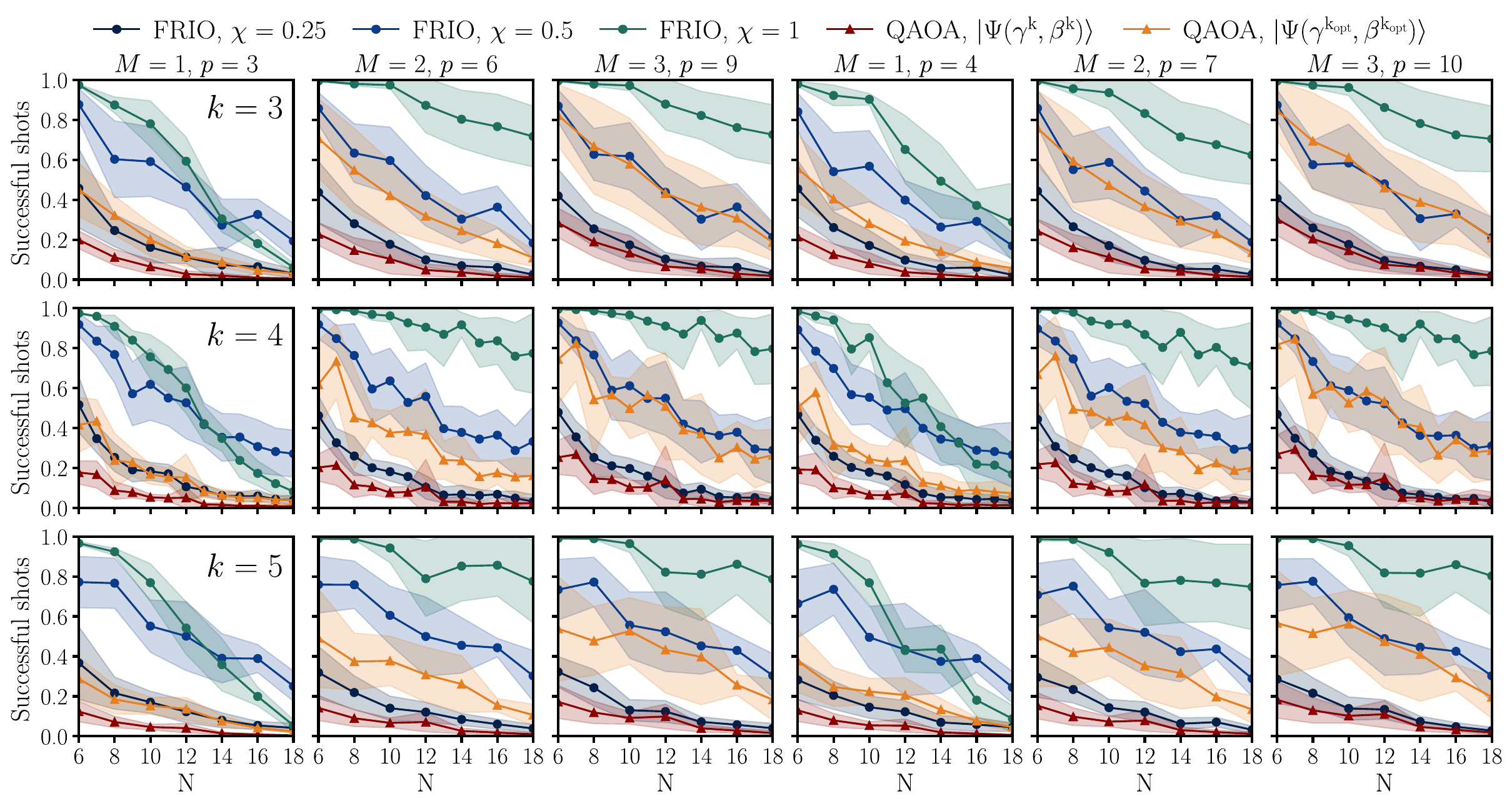}
    \caption{Fraction of shots that successfully solve the max-cut problem for an instance of a weighted $k$-regular graph using digitised FRIO and a \gls{qaoa} circuit of the same depth. The top, middle and bottom rows correspond to a weighted $k=3$, $k=4$, and $k=5$ regular graph, respectively. Each column -- in order of left to right -- corresponds to a first-order Trotter decomposition with $M=1,2,3$ Trotter steps, then a second-order Trotter decomposition with $M=1,2,3$ Trotter steps. In each column we compare to a \gls{qaoa} circuit of depth $p=3,6,9,4,7,10$, respectively. For the cyclic optimisation algorithm we choose $\chi=0.25,0.5,1$ and randomly select 50 starting reference states and allocate 1000 shots to each of them. For \gls{qaoa} we take $5\times10^4$ samples for each graph instance using the known angles from~\cite{Shaydulin2023skatan,Sureshbabu2024skatan,cepaite2025QAOAwarmstarts} and the angles after optimizing for each graph instances. All simulations have been performed for graphs with $N$ between 6 and 18 nodes and averaged over 10 graph instances with the semi-transparent shaded areas representing the statistical error.}
    \label{fig:dreg_shot_resources}
\end{figure*}
Since we have shown that our parameter testing algorithm is sufficient to detect chaotic phases in $k$-regular spin glass Hamiltonians, we now demonstrate that it is possible to apply our $\mathrm{FRIO}$ algorithm to the same models. Just as we do in the main text, we want to benchmark our algorithm against \gls{qaoa} using the average and optimised angles $(\gamma^{k},\beta^{k})$ and $(\gamma^{k_{\text{opt}}},\beta^{k_\text{opt}})$, respectively. Our results are shown in \cref{fig:dreg_shot_resources} where we benchmark our digitised cyclic algorithm for $\chi=0.25,0.5,1$ against the two cases of \gls{qaoa} for $10$ graph instances between $6$ and $18$ nodes. Just like in the main text, we randomly select $50$ starting reference states and allocate $1000$ shots for each of them while for \gls{qaoa} we take $5\times10^4$ samples for each graph instance. Each column -- in order of left to right --  corresponds to a first-order Trotter decomposition with $M=1,2,3$ Trotter steps, then a second-order Trotter decomposition with $M=1,2,3$ Trotter steps. They are compared to a \gls{qaoa} circuit of depth $p=3,6,9,4,7,10$ respectively. Finally, we set $B_z=1$ for ease.

We begin by analyzing the top row of \cref{fig:dreg_shot_resources}, which shows the comparison between \gls{qaoa} and $\mathrm{FRIO}$ for $k=3$ regular spin glass Hamiltonians. We observe that for $\chi=1$ and all considered choices of Trotter decomposition our FRIO algorithm consistently beats a fully optimized \gls{qaoa} circuit for all considered system sizes. Just as we observed for the \gls{sk} model in the main text, for $M=1$ Trotter step we see a sharp decrease in the ability of $\mathrm{FRIO}$ to find the ground state due to the Trotterisation becoming less effective for larger $N$. This breakdown allows for more Landau-Zener transitions to higher energies causing our algorithm to become stuck in mid spectrum eigenstates. For $\chi=0.5$ our algorithm also notably beats optimized \gls{qaoa} for a small number of Trotter steps or it is comparable to optimized \gls{qaoa} for $M=3$ Trotter steps. This can be further seen when looking at the statistical error -- the shaded regions on  \cref{fig:dreg_shot_resources} -- which overlap considerably, demonstrating that this phenomenon is consistent across multiple graph instances at the same size. It is also worth noting that $\chi=0.25$, which does not lie far into the chaotic regime of the phase diagram of \cref{eq:PDHamiltonianklocal}, also remains competitive with the \gls{qaoa} wavefunction from the angles $(\gamma^{k},\beta^{k})$ even for larger graph instances. 

Next we consider weighted $k=4$ regular Hamiltonians which we show in the middle row of \cref{fig:dreg_shot_resources}. This shows a similar performance as demonstrated for $k=3$ regular graphs. As an example, $\chi=1$ consistently outperforms fully optimized \gls{qaoa} for all considered Trotter decompositions but shows a dramatic decrease in efficacy at larger graph instances for $M=1$ Trotter step. Moreover, $\chi=0.5$ also either outperforms or remain competitive with fully optimized \gls{qaoa} even for the largest considered graph sizes. For the final considered quench parameter, $\chi=0.25$, we also observe that our digitised algorithm either outperforms or matches the performance of the \gls{qaoa} wavefunction obtained using the angles from the \emph{\bf{SKatan}} method. Finally, we obtain consistent behaviour with $k=5$ regular Hamiltonians for all considered values of $\chi$ as shown in the bottom row of \cref{fig:dreg_shot_resources}. Importantly, this highlights the broad range of spin glass Hamiltonians that our algorithm can be applied to and how important it is to choose the correct parameters for the cyclic quench sequence. Finally, it is worth highlighting that the \emph{\bf{SKatan}} method is a poor choice as a parameter setting technique for $k$-regular graphs and that optimisation of the angles is generally required. For classically intractable graph instances, this would require to perform the optimisation procedure on quantum hardware. Just like for the \gls{sk} model, it would also be interesting to see whether our algorithm could outperform other parameter setting warm-start techniques like using machine learning and graph neural networks \cite{Deshpande2022GNN,Liang2024GNN} which we leave to future investigations.

\section{Level statistics analysis}\label{sec:level stats}

\begin{figure}
    \centering
    \includegraphics[width=0.9\linewidth]{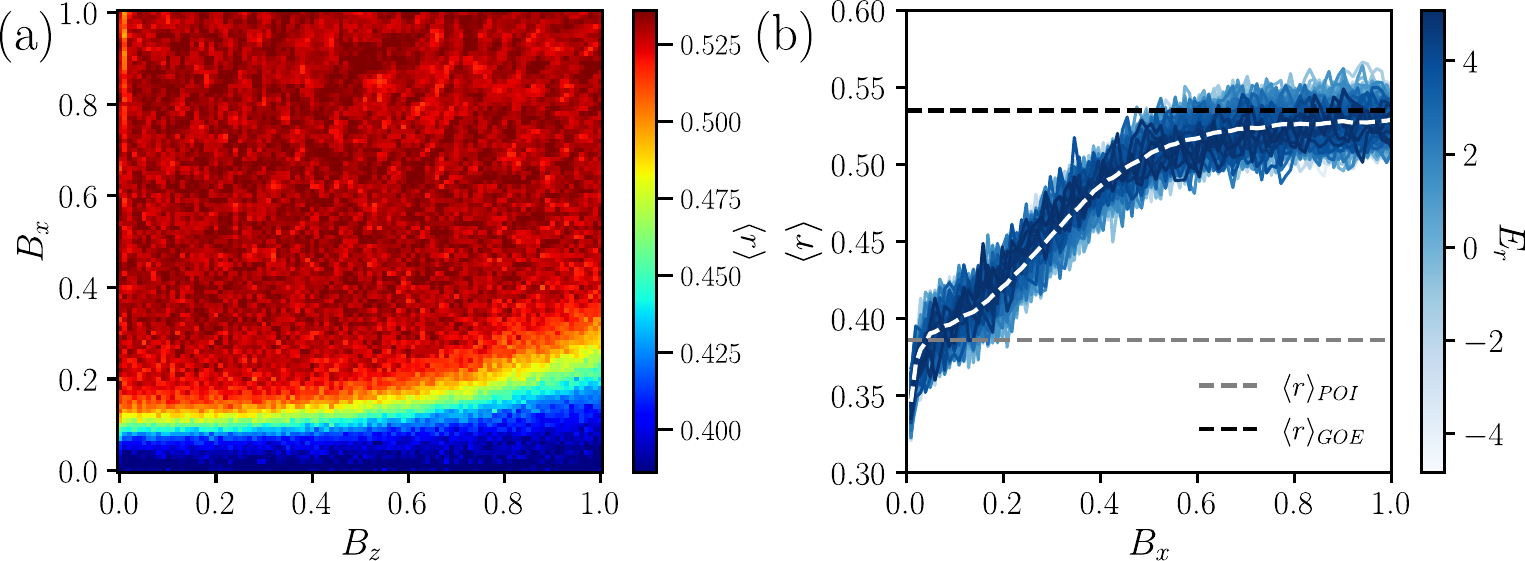}
    \caption{Level statistics analysis of the phase diagram of the \gls{sk} model. (a): Approximate phase diagram of the \gls{sk} model obtained by the average repulsion in the energy spectrum as defined in \cref{eq:r metric}. The data is obtained for a single instance of an $N=12$ node \gls{sk} graph and a single reference state $S^r$. (b): Averaged level spacing ratios of all $2^N$ possible reference states across the $B_z=1$ cut in the full phase diagram with the average over all $S^r$ given by the dashed white lines. Data is for a single $N=10$ node \gls{sk} graph and we colour each line by the energy of the reference state. The horizontal grey and black dashed lines are $\langle r\rangle_{\rm POI} \approx 0.386$ and $\langle r\rangle_{\rm GOE} \approx 0.536$.}
    \label{fig:level_stats}
\end{figure}

As an initial diagnostic to determine whether the chosen Hamiltonian parameters lie in the thermal phase we choose to examine the level-spacing statistics. Spectral diagnostics are typical tools used to study both integrable and chaotic quantum systems \cite{Oganesyan2007levelstats}. Moreover, integrable and chaotic quantum systems are characterized by distinct spectral statistics \cite{mehta2004randommatrices}, which can be used to discern localized and thermal phases. For our analysis the relevant universality class is that of random real hermitian matrices, described by the Gaussian Orthogonal Ensemble (GOE). We will now introduce this diagnostic  as a method to understand the phase diagram of the \gls{sk} model.

For a random hermitian matrix $H$ with an eigenvalue spectrum $\{E_{n} \}$, where $E_n$ are ordered increasingly, the consecutive level spacings are defined as

\begin{equation}\label{eq:level_space}
    s_{n}=E_{n+1} - E_{n},
\end{equation}
and the ratio of adjacent spacings as:

\begin{equation}\label{eq:r metric}
    r_{n}=\frac{\text{min}(s_n,s_{n-1})}{\text{max}(s_n,s_{n-1})}.
\end{equation}
The statistics of $r_{n}$ is a basis-independent diagnostic of spectral correlations and thus can be used as a detector of chaotic phases. The level-spacing statistics of localized systems should follow a Poisson distribution

\begin{equation}\label{eq:Poisson}
    P_{\text{POI}}(r)=\frac{2}{(1+r)^2},
\end{equation}
with an average gap ratio of $\langle r \rangle_{\text{POI}}\approx0.386$. Notably, the distribution is peaked at $r=0$, indicating the absence of level repulsion. For chaotic phases, it is described by the GOE distribution \cite{Alessio2014levelstatsfloquet}

\begin{equation}
    P_{\text{GOE}}(r)=\frac{27}{4} \frac{r+r^2}{(1+r+r^2)^{5/2}},
\end{equation}
with a corresponding mean gap ratio of $\langle r \rangle_{\text{GOE}}\approx0.536$. Moreover, the distribution vanishes as $r\rightarrow0$ indicating that there is level repulsion. In practice, when numerically diagonalizing Hamiltonians, we must resolve all symmetries to remove energy degeneracies which can push chaotic spectra towards a Poissonian distribution.

In \cref{fig:level_stats}(a), we use the average level spacings ratio $\langle r\rangle$ as a proxy to understand the phase diagram of the \gls{sk} model. We consider a single \gls{sk} graph instance with $N=12$ nodes and a single reference state $S^{r}$. Along the $B_z=0$ cut the \gls{sk} model contains a $\mathbb{Z}_2$ symmetry so we restrict to the $z=+1$ spin-inversion symmetry sector whereas the rest of the phase diagram has no symmetries due to the symmetry breaking longitudinal field. We observe two distinct phases, a MBL phase characterized by Poissonian level statistics and a thermal phase which is accurately described by the GOE distribution. Importantly, this allows for a rough identification of the critical line separating the localized and delocalized phases. At $B_z=1$ we find that the critical point lies at $B_x \approx 0.25$ suggesting that our parameter choices of $\chi=0.5,1.0$ are acceptable choices for performing cyclic optimisation.

Now that we have a rough location of the MBL-thermal phase transition, we need to determine whether the energy density of the reference state can alter the location of the phase transition, potentially requiring a quench further into the thermal regime as we traverse further down the spin glass energy landscape. In \cref{fig:level_stats}(b) we calculate the level statistics of 1 instance of an $N=10$ \gls{sk} graph along the $B_z=1$ cut for all the possible $2^N$ reference states. We colour each line corresponding to the location of the reference state in the energy landscape for the considered \gls{sk} instance and highlight the mean level statistics over all the reference states by the white dashed line. Importantly, we observe that the location of the phase transition does not change depending on the chosen reference state. Therefore once the quench parameters are chosen, they can be fixed throughout the entire optimisation protocol.

\section{Other thermalization measures and basis measurements}\label{sec:measurement bases}

For detecting the localisation to delocalized phase transition, we introduced a parameter testing algorithm based on short-time \gls{ipr} dynamics which can be implemented directly on quantum hardware. However, we kept the analysis confined to measurements performed in the computational basis, which may not be indicative of complete thermalization in all possible bases. Therefore, we now observe the effect that measurement basis has on the efficacy of our parameter selection algorithm alongside other measures for thermalization other than the considered $4$-qubit marginal.

\begin{figure}
    \centering
    \includegraphics[width=0.99\linewidth]{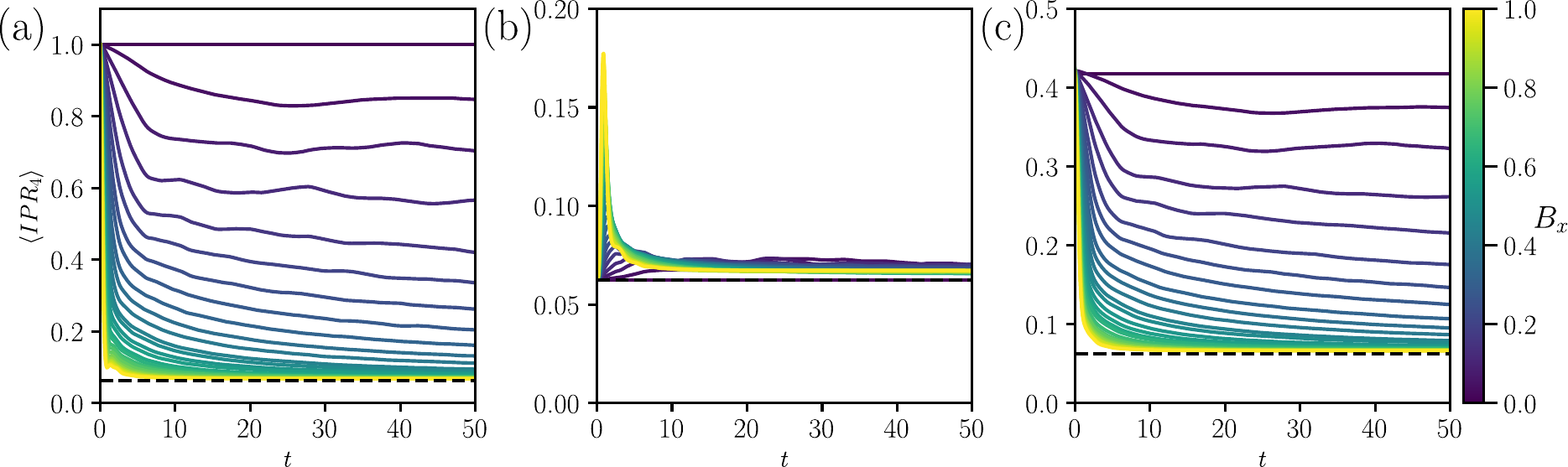}
    \caption{Marginal \gls{ipr} dynamics for the \gls{sk} model using measurements in different bases using exact diagonalization. (a): The long-time \gls{ipr} dynamics of a $4$-qubit marginal, measured in the $Z$-basis for multiple values of $B_x$ which is indicated by the colour bar. (b): The same as (a) but the \gls{ipr} measurements are performed in the $X$-basis. (c): Also the same as (a) but measured in $20$ random single qubit bases. The dashed black line is the expected thermal value, $\langle \mathrm{IPR}_4\rangle_{th}=1/2^{4}$  All simulations were averaged over $10$ different \gls{sk} instances with $N=14$ nodes and also averaged over $100$ randomly selected reference states $S^{r}$. The initial state is the product state $| \psi(0)\rangle=|01...01\rangle$.}
    \label{fig:IPR dynamics bases} 
\end{figure}

In \cref{fig:IPR dynamics bases}(a) we plot the $4$-qubit \gls{ipr} dynamics measured in the $Z$-basis, for multiple values of the transverse field strength $B_x$ using the initial state $| \psi(0)\rangle=|01...01\rangle$. We choose to average over $10$ instances of an $N=14$ \gls{sk} graph and $100$ randomly selected initial reference states. As observed in the main text, we see that the $Z$-basis is an excellent basis to choose, with the dynamics converging to the expected thermal value of $\langle \mathrm{IPR}_4\rangle_{th}=1/2^{4}$ when the selected Hamiltonian parameters are located in the thermal phase; see \cref{sec:results} in the main text for a more detailed discussion. On the other hand, \cref{fig:IPR dynamics bases}(b) performs the same analysis where the \gls{ipr} measurements are performed in $X$-basis. Unfortunately, in this basis we observe delocalized dynamics for all considered values of $B_x$, indicating that the $X$-basis is not a good measurement basis for our parameter selection algorithm. The behaviour can be explained as $|\psi(0)\rangle$ is already delocalized in the $X$-basis, whereas the localisation physics is contained in the \gls{sk} and longitudinal field terms. Therefore, any delocalisation caused by the transverse field term will remain undetected when measuring in the $X$-basis. However, our parameter selection protocol remains robust when considering random single qubit measurements as demonstrated in \cref{fig:IPR dynamics bases}(c), where we also average over $20$ randomly selected single qubit bases. Here we are able to see a clear difference in the localized and the delocalized dynamics where the delocalized dynamics converge to the expected thermal value.

\begin{figure}
    \centering
    \includegraphics[width=0.75\linewidth]{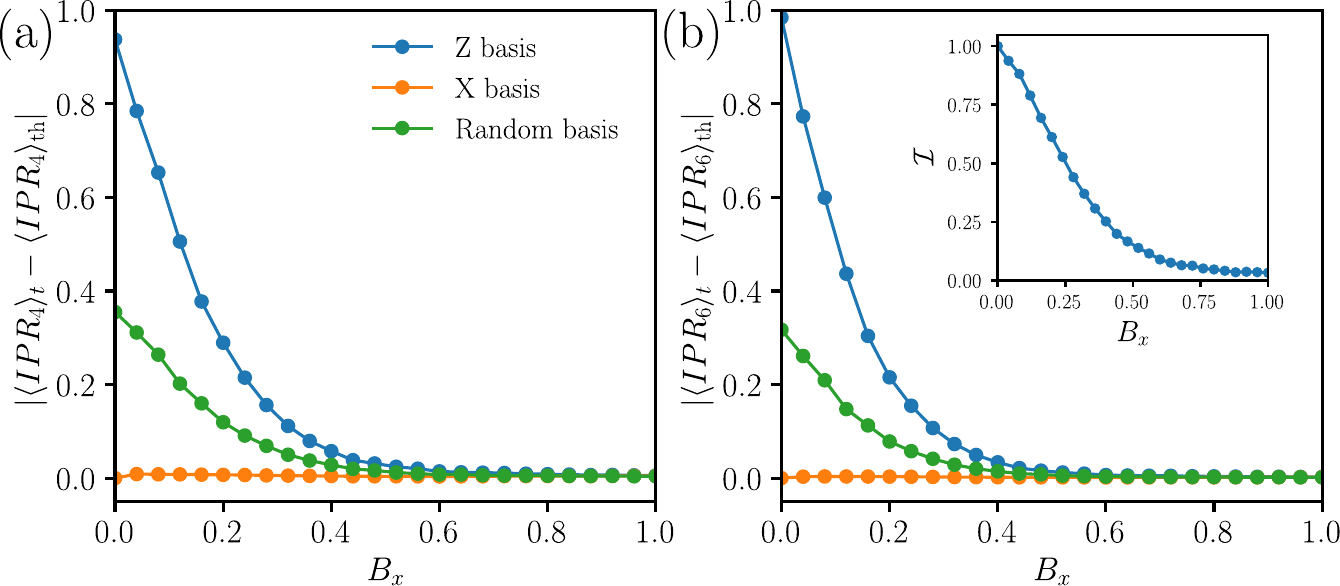}
    \caption{Detecting the MBL-thermal phase transition of \gls{sk} graphs using different thermalization measures. (a) Late time average of the \gls{ipr} dynamics for a $4$-qubit marginal shifted by the expected thermal value of the \gls{ipr} located at multiple transverse field strengths $B_x$. (b): The same as (a) but for a $6$-qubit marginal. Here, $\langle \mathrm{IPR}_A\rangle_{th}=1/2^{|A|}$, the expected thermal value of the $|A|$ qubit subsystem. Inset: The late time average of the spin-imbalance given by~\cref{eq:imbalance} dynamics as a function of transverse field strength $B_x$. All the simulations were performed using the \gls{sk} instances and reference states from~\cref{fig:IPR dynamics bases} and the late-time averages have been calculated between $t=30$ and $t=50$.}
    \label{fig:IPR QPT}
\end{figure}

We also consider the effect that the chosen measurement basis has on the ability of the Hamiltonian to thermalize using long-time dynamics. In \cref{fig:IPR QPT}(a) we show $\langle \mathrm{IPR}_4\rangle_t$, the time-averaged \gls{ipr} between times $t=30$ to $t=50$ for the dynamics shown in \cref{fig:IPR dynamics bases}. Like in the main text, we plot the norm of the difference between the time-averaged $\langle \mathrm{IPR}_4\rangle_t$ and the expected thermal value $\langle \mathrm{IPR}_4\rangle_{th}$. As we observed for the dynamics in \cref{fig:IPR dynamics bases}, we are easily able to discern the localized and delocalized phases when the measurements are performed in the $Z$ or a randomly selected single qubit basis. The $X$-basis measurement remains fully delocalized for all values of $B_x$ rendering it a poor choice for this initial state. These trends continue when we change our thermalization metric to a $6$-qubit marginal, which is plotted in \cref{fig:IPR QPT}(b). Here we observe that the choice of the size of the marginal has a negligible effect on the late-time dynamics, with the behaviour barely changing for the three different basis measurements. Therefore, this adds additional confirmation that the $4$-qubit marginal is a satisfactory thermalization measure for our parameter testing algorithm.

Finally, we also consider a more traditional, dynamical measure of thermalization, called the spin-imbalance,
\begin{equation}\label{eq:imbalance}
    \mathcal{I}(t)=\frac{1}{N}\sum_{j=1}^{N}(-1)^{j-1}\left\langle Z_j(t) \right\rangle,
\end{equation}
which is a global measure of thermalization compared to the previous local measures. In the inset of \cref{fig:IPR QPT}(b) we plot the time averaged spin imbalance for the same \gls{sk} graph instances and reference states as we cross the localized to thermal critical line at $B_z=1$. In particular, we perform the averaging between times $t=30$ and $t=50$. While the spin-imbalance is able to discern the two phases, the range in which the location of the critical line could be located is much larger compared to the time averaged \gls{ipr}. Therefore, the spin-imbalance could also be a reasonable candidate for our parameter testing algorithm with the added caveat that it requires measurements on all sites compared to measurements over a small subsystem. We do note that it is possible to consider a spin-imbalance analog which is defined over a marginal of size $A$, similar to our previous \gls{ipr} analysis.

\section{Finite-size scaling using successful energy transitions}\label{sec:succes fss}

\begin{figure}[h]
    \centering
    \includegraphics[width=0.97\linewidth]{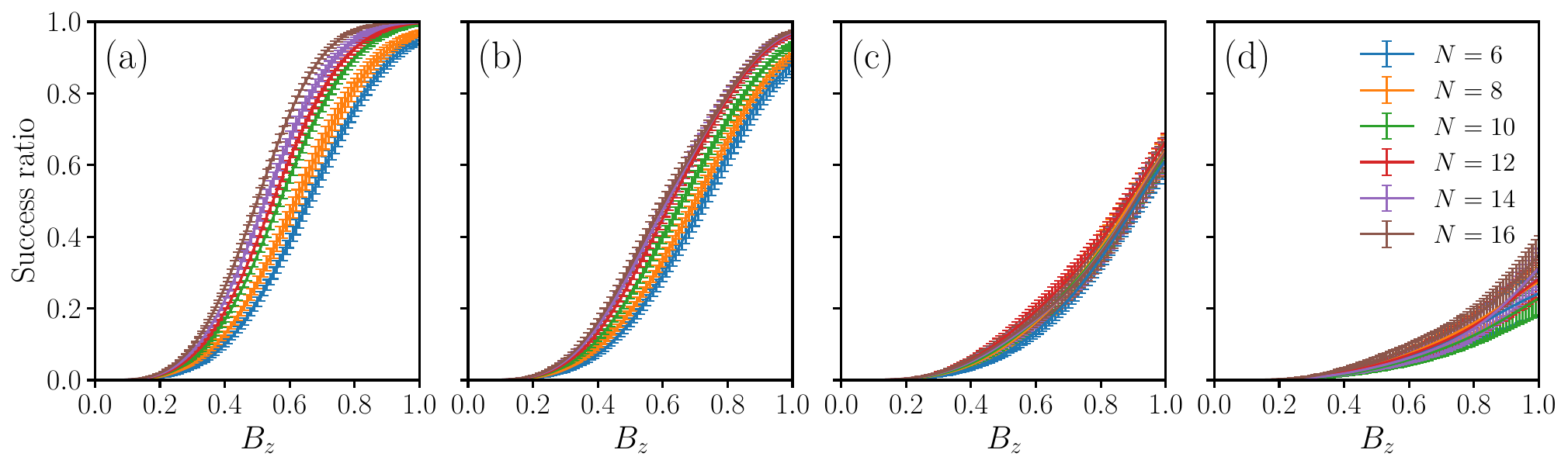}
    \caption{Finite-size scaling of the success ratio at multiple energy densities for the SK model. The panels (a)-(d) correspond to the finite-size scaling of the success ratio where the initial reference state is located the closest to the energy density $\varepsilon=(E-E_{\text{min}})/(E_{\text{max}}-E_{\text{min}})$ of $\varepsilon=0.8,0.5,0.2,0.05$ respectively. The error bars represent the statistical error of averaging over 10 distinct SK graph instances. We also set $B_x=\chi B_z$, with $\chi=1$, and the chosen quench times are $\tau_1=1$, $\tau_2=\chi B_z$, and $\tau_3=\chi B_z$.}
    \label{fig:Success_ratio_fss}
\end{figure}

During the initialisation of the FRIO algorithm, there are many possible parameters that are set before the procedure is executed. These parameters include the field strengths $B_z$ and  $\chi$,  the quench durations $\tau_1, \tau_2, \tau_3$, and the initial reference state $S^{r}$ which we further analyze in \cref{sec:reference state analysis}. However, it is important to gauge what is the likelihood that the algorithm will eventually traverse the energy landscape and find the ground state instead of becoming stuck at local minima. Most likely, this will happen as the reference state reaches lower energy densities, especially at larger graph sizes, where we expect that a larger number of Trotter steps $M$ will be required for FRIO to be able to get closer to the ground state. Therefore, in this section we show how system size affects the algorithm's ability to transition to lower energies at multiple points in the \gls{sk} eigenspectrum. 

As a measure for the probability of a successful energy transition after performing one cycle for a given reference state, we define the success ratio (equivalent to \cref{eq:succ_prob} in the main text):

\begin{equation}\label{eq:success ratio}
    \text{Success ratio}=\sum_{i: E_{i }<E^{r}}p_i, 
\end{equation}
where $E^{r}$ is the energy of the initial reference state, $p_{i}$ is the probability of measuring a bitstring $i$, and $E_{i}$ is that bitstring's energy. For small systems that can be simulated using ED this calculation can be performed exactly but for larger systems where state sampling will be required we can choose an alternative definition:

\begin{equation}
    \text{Success ratio}=\frac{N_{E_{i}<E^{r}}}{N_{\text{shots}}},
\end{equation}
where $N_{\text{shots}}$ is the total number of shots used during a single cycle of the algorithm and $N_{E_{i}<E^{r}}$ is the number of measured bitstrings which are located lower in the energy landscape compared to the original reference state $S^{r}$. This becomes an incredibly useful metric to characterize successful energy transitions during a single cycle as the energy of the measured bitstrings can be easily calculated for large system sizes. We choose to limit ourselves to system sizes where the success ratio can be calculated exactly to remove any sampling errors. We perform finite-size scaling analysis for \gls{sk} graphs that contain between $6$ and $16$ nodes and average over 10 realisations of  \gls{sk} graphs for each system size. For each \gls{sk} instance we choose the reference state that lies the closest to some chosen energy density $\varepsilon$ and perform a single quench cycle with the quench rates used in the main text.

In \cref{fig:Success_ratio_fss}(a) we plot finite-size scaling of the success ratio at $\chi=1$ as a function of $B_z$ where the initial reference state is located at an energy density $\varepsilon\approx 0.8$. The error bars represent the statistical error after averaging over the 10 \gls{sk} realisations. Here we observe excellent finite-size scaling where the success probability actually increases as we increase system size. This is due to the number of reference states that lie lower in the energy landscape compared to the original state, increasing the probability of a successful energy transition. Most notably, at $B_z=1$ and $\chi=1$ for all $N\gtrsim 10$ we are highly likely to achieve a successful energy transition for this choice of quench sequence. A similar conclusion can be ascertained when the initial reference state is located at $\varepsilon=0.5$ which we plot in \cref{fig:Success_ratio_fss}(b). At this energy density we still obtain good finite-size scaling where the success ratio still increases as we increase system size though not as much. Since we are located directly in the middle of the eigenstate spectrum, the number of eigenstates that are at lower energies is much lower decreasing the overall probability of a successful energy transition though this does not seem to have that powerful of an effect.

When finding the ground state of the \gls{sk} model using our FRIO algorithm it is expected to rapidly progress through the reference states located at high energies before slowing down at low lying excited eigenstates. In particular, it is possible for our algorithm to become stuck in low energy eigenstates that are not necessarily the ground state.  To demonstrate that this is unlikely, we also perform finite-size scaling in the success ratio for reference states that are low lying energy eigenstates. In \cref{fig:Success_ratio_fss}(c) we perform the same analysis for reference states at energy density $\varepsilon=0.2$ where we also observe excellent finite-size scaling with the curves corresponding to all the considered system sizes converging. Importantly, for the parameters considered in the main text $B_z=1$ and $\chi=1$, there is $\approx60\%$ chance that we obtain a successful energy transition where we measure a lower energy bitstring. Finally, we perform the same analysis for reference states located extremely close to ground state, $\varepsilon=0.05$, which we plot in  \cref{fig:Success_ratio_fss}(d). Here we observe mostly converged success ratio curves for all system sizes across most values of $B_z$, with an $\approx30\%$ chance of a successful energy transition. We also note that for the two low-energy densities considered here, it does not appear that the finite-size scaling could approach zero in the infinite-size limit, demonstrating that our algorithm should retain a significant success probability over a wide range of graph sizes once we scale beyond classically tractable graph instances.

\section{The effect of Trotter order and decomposition} \label{sec:Trotter order}

\begin{figure}[h]
    \centering
    \includegraphics[width=0.94\linewidth]{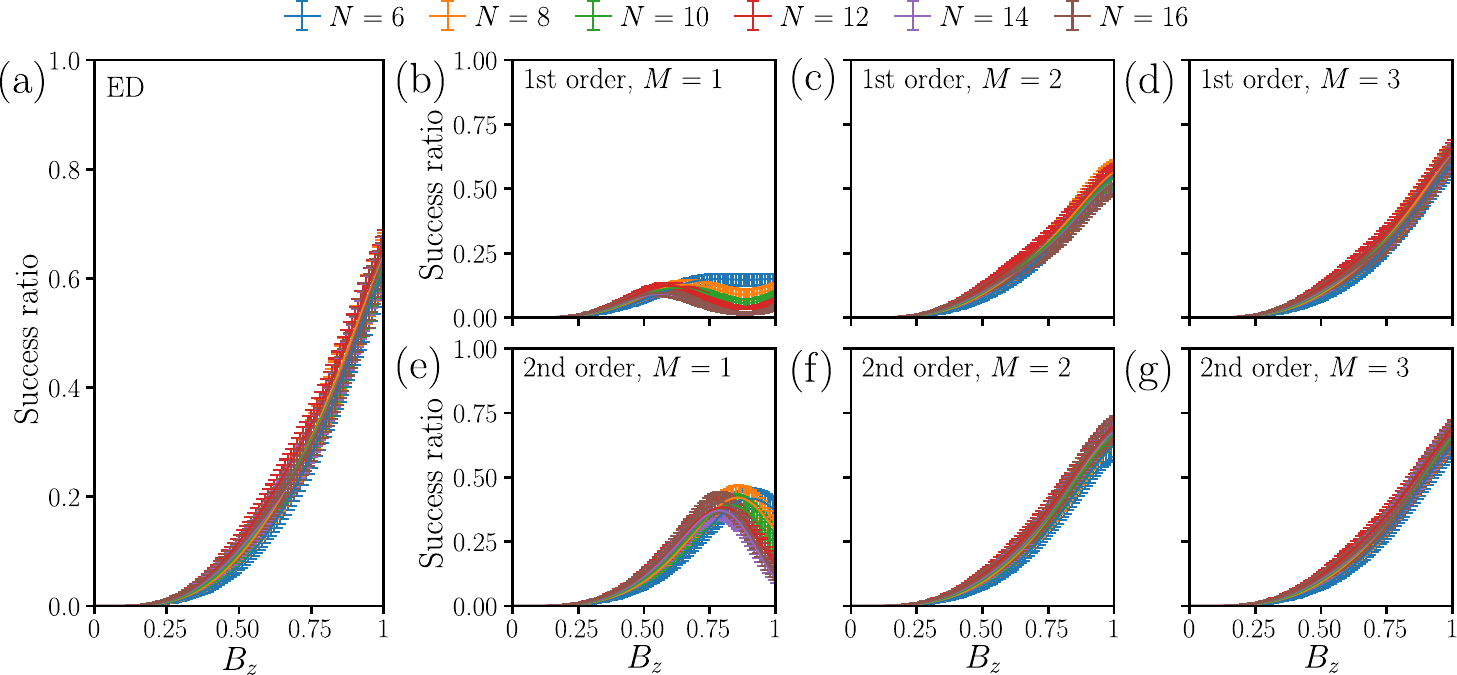}
    \caption{The effect on finite-size scaling due to Trotterisation on the probability of a successful energy transition. (a): finite-size scaling of the success ratio using continuous time evolution. The panels (b)-(d) correspond to the finite-size scaling of the digitised protocol using a first-order Trotter decomposition and $M=1,2,3$ Trotter steps. The panels (e)-(g) contains the same analysis instead using a second-order Trotter decomposition and $M=1,2,3$ Trotter steps. The error bars represent the statistical error from averaging over $10$ distinct SK graph instances with the closest initial reference state to the energy density $\varepsilon=0.2$. We set $B_x=\chi B_z$ where $\chi=1$ and the chosen quench times are $\tau_1=1$, $\tau_2=\chi B_z$ and $\tau_3=\chi B_z$.}
    \label{fig:Trotter fssa}
\end{figure}

In the previous section we demonstrated, using finite-size scaling techniques, that the chosen quench trajectory has a low likelihood of becoming stuck in low-lying excited eigenstates using continuous dynamical simulations. However, since our digital protocol requires Trotterisation to run on hardware, the low number of Trotter steps we consider in the main text may alter the behaviour of the protocol, reducing the chances for the algorithm to reach the ground state. We now demonstrate that for the chosen non-adiabatic quench times, $\tau_1=1$, $\tau_2=\chi B_z$ and $\tau_3=\chi B_z$, the use of a Trotter decomposition does not have a notable adverse effect on probability of the algorithm undergoing a successful energy transition.

Just as we did in \cref{sec:succes fss}, we perform a finite-size scaling of the success ratio,~\cref{eq:success ratio}, for classically tractable graphs after Trotterisation. We average over 10 different \gls{sk} instances and initialise the quench protocol with the reference state at energy density $\varepsilon=0.2$. We also use the continuous dynamics where we set $\chi=1$ shown in \cref{fig:Trotter fssa}(a) as a benchmark. In \cref{fig:Trotter fssa}(b)-(d) we perform the same finite-size scaling in the success ratio for our FRIO algorithm using a first-order Trotter decomposition and $M=1,2,3$ Trotter steps respectively. While $M=2$ and $M=3$ have sufficiently converged to the continuous dynamics, we see a significant breakdown for $M=1$. We observe that at $B_z=1$ the chance of a successful energy transition decreases massively as we increase the size of the \gls{sk} graph. This behaviour causes the large decrease in the performance of our algorithm for large $N$ in~\cref{fig:SK_shot_resources}(a) of the main text.

Similarly, in~\cref{fig:Trotter fssa}(e)-(g) we perform the same analysis but using a second-order Trotterised circuit and $M=1,2,3$ Trotter steps. Just like the first-order decomposition for $M=2$ and $M=3$, the dynamics have sufficiently converged such that the low depth of the circuit should have no meaningful effect on the algorithm's ability to arrive at the ground state. The $M=1$ decomposition identifies another breakdown in the dynamics but not as severe as the first-order decomposition with there still being a small but not negligible chance of a successful energy transition occurring. This explains the performance decrease of our algorithm at $\chi=1$ for large $N$, see~\cref{fig:SK_shot_resources}(d) of the main text, but demonstrates an overall better performance than the first-order decomposition.

\section{Dependence on the initial reference state} \label{sec:reference state analysis}

In \cref{sec:succes fss} we previously demonstrated that the cyclic trajectory our algorithm traverses scales favourably to larger system sizes no matter where the initial reference state is located in the spin glass energy landscape. Therefore, the quench parameters that we choose when initialising the algorithm can be maintained throughout the optimisation procedure. However, the previous finite-size scaling was limited to reference states located at fixed energy densities across different SK graphs. In this section we look at the explicit dependence on the initial reference state $S^{r}$ of the algorithm's ability to transition to low-energy states.

\begin{figure}
    \centering
    \includegraphics[width=0.85\linewidth]{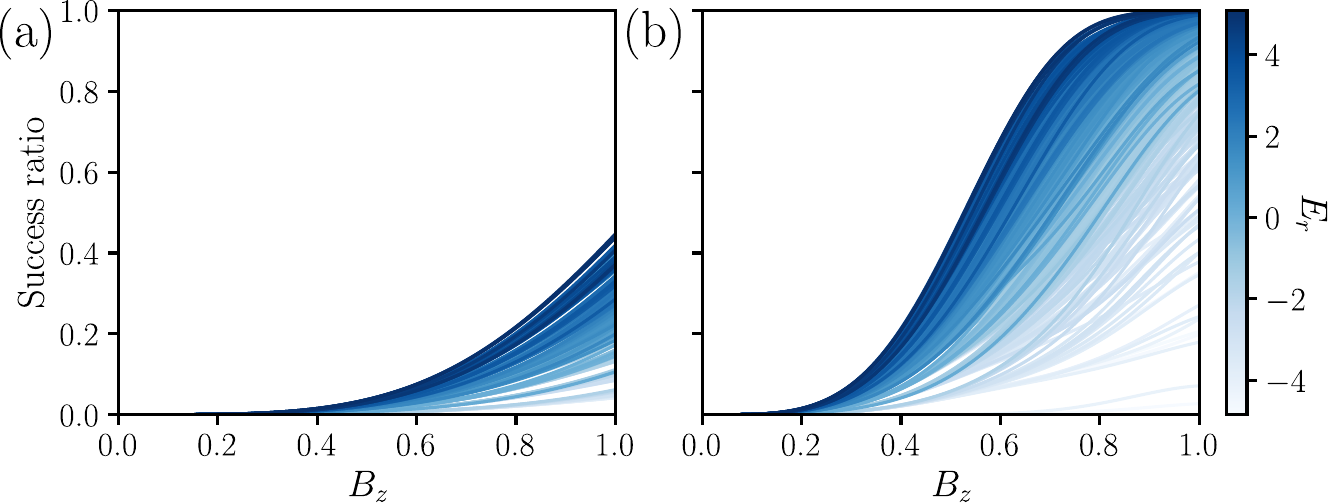}
    \caption{Dependence of the initial reference state when traversing through the glassy energy landscape. (a): Success ratio of all possible starting reference states of a single \gls{sk} graph with $N=10$ nodes for $\chi=0.5$ for a single optimisation cycle. (b): The same as (a) but for $\chi=1$. We colour each line by the location of the reference state in the energy landscape in the \gls{sk} graph with the ground state located at $E_{\rm gs}=-4.845$. We set the quench times to be $\tau_1 =1$, $\tau_2 =\chi B_z$ and $\tau_3=\chi B_z$. In all panels, $B_z = 1$.}
    \label{fig:all states}
\end{figure}

In \cref{fig:all states}(a) we plot the ratio of successful energy transitions as defined in \cref{eq:success ratio} after performing a single optimisation cycle for all $2^N$ possible initial reference states for a single instance of an $N=10$ node \gls{sk} graph using $\chi=0.5$. We colour each line on the plot by the location of the reference in the energy landscape of the \gls{sk} graph. As the energy of the chosen reference state decreases we observe a decrease in the probability of a successful energy transition after one cycle. This is because the number of states below the energy of the initial reference state decreases. This is further shown in \cref{fig:all states}(b) which performs the same analysis but for $\chi=1$, which also demonstrates that the probability of a successful transition to a lower energy state is dependent on the location of the initial reference state in the spin-glass energy landscape. Moreover, it further demonstrates the power of quenching deeper into the thermal phase with some reference states that are located at higher energies have an almost 100\% chance of a successful energy transition.

\end{document}